\documentclass[transmag]{IEEEtran}
\usepackage{latexsym}
\usepackage{amsfonts,amssymb,amsmath}
\usepackage{hyperref}
\def\BibTeX{{\rm B\kern-.05em{\sc i\kern-.025em b}\kern-.08em T\kern-.1667em\lower.7ex\hbox{E}\kern-.125emX}}
\usepackage{amsmath}
\usepackage{cite}
\usepackage{float}
\usepackage{dirtytalk}
\usepackage{array}
\usepackage[shortlabels]{enumitem}
\usepackage{amsthm}
\usepackage[pdftex]{graphicx}

\usepackage[ruled,linesnumbered]{algorithm2e}
\newtheorem{theorem}{Theorem}[section]
\newtheorem{corollary}{Corollary}[theorem]

\usepackage{amssymb}
\usepackage{wasysym}
\usepackage{pifont}
\usepackage{multirow,bigdelim}
\usepackage[makeroom]{cancel}
\usepackage{color}
\usepackage{bm} 
\usepackage{makecell}
\usepackage{authblk}
\def\BibTeX{{\rm B\kern-.05em{\sc i\kern-.025em b}\kern-.08em
    T\kern-.1667em\lower.7ex\hbox{E}\kern-.125emX}}
\usepackage{diagbox}
\usepackage{tikz}
\usetikzlibrary{automata, positioning, arrows.meta,arrows}
\tikzset{
->, 
>=Latex, 
node distance=5cm, 
every state/.style={thick, fill=gray!16}, 
}

\begin{document}

\title{Bayesian Detection of a Sinusoidal Signal with Randomly Varying Frequency}


\author{Changrong Liu, S. Suvorova, R. J. Evans, \IEEEmembership{Life Fellow, IEEE}, B. Moran,  \IEEEmembership{Member, IEEE},  A. Melatos
\thanks{The authors acknowledge support from the Australian Research Council (ARC) through the Centre of Excellence for Gravitational Wave Discovery (OzGrav) (grant number CE170100004) and an ARC Discovery Project (grant number DP170103625).}
\thanks{
Changrong Liu is with the Department of Electrical and Electronic Engineering, University of Melbourne, Parkville, Victoria 3010, Australia (e-mail: changrongl1@student.unimelb.edu.au).}
\thanks{S. Suvorova is with the Department of Electrical and Electronic Engineering, University of Melbourne, Parkville, Victoria 3010, Australia and Australian Research Council Centre of Excellence for Gravitational Wave Discovery (OzGrav) (e-mail: sofia.suvorova@unimelb.edu.au).}
\thanks{R. J. Evans is with 
the Department of Electrical and Electronic Engineering, University of Melbourne, Parkville, Victoria 3010, Australia and Australian Research Council Centre of Excellence for Gravitational Wave Discovery (OzGrav) (e-mail: robinje@unimelb.edu.au).}
\thanks{B. Moran is with 
the Department of Electrical and Electronic Engineering, University of Melbourne, Parkville, Victoria 3010 (e-mail: wmoran@unimelb.edu.au).}
\thanks{A. Melatos is with the School of Physics, University of Melbourne, Parkville, Victoria 3010, Australia and Australian Research Council Centre of Excellence for Gravitational Wave Discovery (OzGrav) (e-mail: amelatos@unimelb.edu.au).}}

\IEEEtitleabstractindextext {\begin{abstract}
The problem of detecting a sinusoidal signal  with randomly varying frequency has a long history. It is one of the core problems in signal processing, arising in many applications including, for example, underwater acoustic frequency line tracking, demodulation of FM radio communications, laser phase drift in optical communications and, recently, continuous gravitational wave astronomy. In this paper we describe  a Markov Chain Monte Carlo based procedure to compute a specific detection posterior density. We  demonstrate via simulation that our approach results in an  up to  $25$ percent higher detection rate than   Hidden Markov Model based solutions, which are generally considered to be the leading techniques for these problems.
\end{abstract}

}

\maketitle

\section{Introduction}
\IEEEPARstart{T}{\lowercase{h}}e problem of  detecting a sinusoidal
signal with randomly varying frequency,  measured in additive noise,  is encountered in numerous applications.  Our interest  derives from attempts to detect the presence of as yet undiscovered gravitational waves hypothesized to emanate from rotating astronomical objects like neutron stars \cite{Jaranowski_1998}, \cite{Riles_2013}, whose frequency wanders slowly and randomly \cite{10.1093/mnras/stt1828}. Attempts to develop optimal and good sub-optimal
solutions have occupied  many signal processing
researchers for at least 50 years, including more recent work in \cite{janvcovivc2011detection, comar2021detection, TUAN2019245, 8897612}. In essence, the problem can be categorized as detection of a non-Gaussian random process in Gaussian noise, and the forms of the optimal detector are well known~\cite{ poor2013introduction,kailath1998detection}. 
%
However, these  require the  conditional-mean estimate of the signal which, apart for a small number of cases, is extremely difficult to characterize and compute.

 Many approximate solutions have been  developed including use of an extended Kalman filter (EKF) to track the random frequency followed by a coherent detector \cite{1096303}, \cite{gardner2005phaselock}. This approach is known to be far from optimal because of the simple linearization used in the EKF. Another commonly used approximation is to assume a quadratic detector structure \cite{1089403} and optimize a relevant performance cost, typically the deflection ratio.  This approach leads to the use of the covariance of the random signal in the quadratic detector, but this is not optimal for a non-Gaussian  random signal. One class of approximate methods relies on Hidden Markov Models (HMMs) and the Viterbi algorithm~\cite{macchi1981dynamic,short1982detection,1056241} to rapidly compute the  maximum a posterior (MAP) estimate using the short-time discrete Fourier transform (DFT). The detector is then formed by substituting in the MAP estimate. These methods rely on a Markov assumption for the wandering frequency dynamics between  time blocks.  An extension of this method enforces  phase continuity between time blocks \cite{suvorova2018phase},  resulting in  a further improvement. Attempts to replace the short-time DFT with more sophisticated time-frequency analysis methods, such as the Wigner-Ville distribution,  appear to offer no performance advantage over a standard Viterbi approach in terms of frequency tracking accuracy or detection performance \cite{djurovic2004algorithm}.

In this paper we form the detection  statistic by directly computing a specified posterior density using Markov Chain Monte Carlo (MCMC) \cite{e2544badcbd543a481cc0bdf45041dc4} methods. The essential idea behind MCMC is to construct a Markov chain of which the invariant distribution is the desired posterior distribution. When the Markov chain converges to its equilibrium, the samples generated by the chain are essentially samples from the posterior distribution of interest \cite{e2544badcbd543a481cc0bdf45041dc4}. To construct the detector, we  introduce a time-invariant binary random variable $k$ that indicates whether a signal is present in the data or not. Thus, under the null hypothesis $H_0,\:(k=0)$, the signal of interest is absent and under the alternative hypothesis $H_1,\:(k=1)$, the signal is present, with unknown amplitude and wandering frequency.  The detection statistic is then set to be the posterior distribution  of $k$, given observations $\mathbf{y}$,  denoted by $\Pr(k|\mathbf{y})$.  This posterior involves two distinct terms,  $\Pr(k=0|\mathbf{y})$ and $\int_{\Theta}\Pr(k=1,\theta|\mathbf{y})d\theta$, with $\theta$ being a parameter in the space $\Theta$ of unknown amplitude, wandering frequency and phase of the signal under $k=1$. 
 We show, in order to evaluate $\Pr(k|\mathbf{y})$, we have to estimate ${\theta}$ as well.
A closely related idea is used in \cite{andrieu1999joint},  where  the signal is modelled as a superposition of several single frequency sinusoids; in that case  the number of sinusoids as well as their corresponding (constant) frequencies is estimated.
We differ from previous work in that we focus on detecting  one wandering frequency line,  modelled as a high dimensional unknown parameter vector. The generalization to multiple  signals is straightforward, albeit at the price of increased computational complexity.

 In this work we follow an important extension to the basic MCMC method, called \emph{reversible jump MCMC} (RJMCMC) \cite{green1995reversible}, which allows samples to jump between multiple spaces with different dimensions while maintaining overall equilibrium. 
We first derive the posterior distribution $\Pr(k|\mathbf{y})$ and proceed  to build  it with RJMCMC.
We then introduce a new method for efficiently proposing a candidate frequency path while maintaining  a reasonable acceptance ratio. 
We develop a parametrized model of frequency dynamics with varying parameter dimension,
where we track the frequency at coarsely spaced time samples (``knots'')  while interpolating between knots with quadratic polynomials. The time  between two adjacent knots is referred to  here as a ``block''. The number of knots in this scheme is equal to  the number of blocks.
 We show significant saving of computational resources with a reduced numbers of knots, sufficient to capture the dynamics of the underlying frequency. This characteristic is valuable since real life  applications usually deal with very large amounts of data
(e.g. observation data of gravitational waves typically involves a scalar amplitude channel sampled $\sim 10^{11}$ times over an observation period lasting one year \cite{Riles_2013}). We also illustrate how to choose the number of blocks for HMM and MCMC respectively. In the end, we perform numerical simulations that demonstrate higher estimation accuracy and detection probability of MCMC, compared with HMM based methods.

The remainder of this paper is organized as follows. In Section~\ref{section: problem statement}, a parameterized signal model is presented. In Section~\ref{sec:HMM}, the HMM-based method is briefly explained. In Section~\ref{sec:derive posterior}, the posterior distribution for detection is formally derived. The complete RJMCMC procedure is developed and described in Section~\ref{section: MCMC}, followed by novel methods of generating a new sample path and producing a proposal sample path for a single MCMC birth and update step, detailed in Algorithms \ref{interp}--\ref{algorithm7}. Numerical results are described in Section~\ref{sec:experiment}, where the detection performance of the algorithm is quantified by receiver operating characteristic (ROC) curves.  The extra information provided by estimating frequency paths  is presented  as part of the detection algorithm, with root mean square error (RMSE) recorded. The  MCMC and the HMM methods are compared in terms of these performance measures.
\section{Problem statement}\label{section: problem statement}
Without loss of generality,  we assume the observed real signals with real additive noise are first  converted into complex signals via the Hilbert transform. Throughout this paper, all derivations and simulations are based on complex  data. Let $\mathbf{y}=\{y(t_n)\}_{n=1,\dots,N}$ be the complex-valued data sequence, observed at $N$ equally spaced instants, $t_1\leq \ldots\leq t_N$, with $t_1=0$ and $T\overset{\triangle}{=}t_{n}-t_{n-1}$ for all $n$. Let $k\in \{0,1\}$ be a statistic constant  during the whole observation period, taking values $0$ or 1, denoting whether the data is composed of pure noise ($k=0$) or signal plus noise ($k=1$).
Observations $\mathbf{y}$, which can have been generated  either under  hypothesis $H_0$ or hypothesis $H_1$, are given by
\begin{subequations}
    \begin{align}
       H_0:y(t_n)&=z(t_n), \; \text{ for } k=0,   \\
        H_1:y(t_n)&=a\exp(2\pi j\phi(t_n)+\psi_0)+z(t_n) \label{1b}\\
        &=\tilde{a}\exp(2\pi j\phi(t_n))+z(t_n),\;\text{ for } k=1,  \label{k=1data}
    \end{align}
\end{subequations}
for $n=1,2,\dots N $, where $\psi_0$ in \eqref{1b}  is the unknown initial phase and, for simplicity, we incorporate it into $\tilde{a}$ in \eqref{k=1data} where $\tilde{a}\overset{\triangle}{=}a\exp(j\psi_0)$ indicates the unknown complex-valued random amplitude.  This allows us to  assume that the phase path starts from $\phi(t_1)=0$. In our context, $\tilde{a}$ and $\psi_0$ are treated as nuisance parameters. The noise, $z(t_n) \sim \mathcal{CN}(0,\sigma^2)$  is distributed as a complex Gaussian
 with variance $\sigma^2$. 
 The unknown signal $\mathbf{y}_c=\{y_c(t_n)\}_{n=1,\dots, N}$ is modelled by
\begin{equation}
\label{yc}
y_c(t_n)=\tilde{a}\exp(2\pi j\phi(t_n)),\;n=1\dots N.
\end{equation}

 
 
Assuming that the continuous time-varying frequency is a Wiener process with zero drift and diffusion constant $\gamma$,  we have
\begin{equation}
    \label{continuous}
    \Bigl\{
    \begin{array}{lr}
       \phi(t) = \int_{0}^t f(s)ds  &  \\
        df(t)=\gamma dB(t) & 
    \end{array},
\end{equation}
where $B(t)$ is the standard Wiener process with $E[B(s)B(t)]=\min(s,t)$. Then the discretized counterpart with sampling interval $T$ is
\begin{equation} 
\label{path}
    \Bigl\{
    \begin{array}{lr}
     \phi(t_{n+1}) = \phi(t_n) + f(t_n) T +w_1(t_n) & \\
       f(t_{n+1}) = f(t_n) + w_2(t_n)  &  \\
    \end{array},
\end{equation}
for $n=1,2\dots,N-1$, with $w_1(t_n)$ and $w_2(t_n)$ representing the  instantaneous phase and frequency noise, which are both zero mean Gaussian random variables.

Introduction of the state variable   $\mathbf{x}(t_n)=[\phi(t_n),f(t_n)]^T$, allows us to write
 ~\eqref{path} in matrix form as
\begin{equation}
\label{origin model}
\mathbf{x}(t_{n+1}) = 
\begin{bmatrix}
     1&T  \\
     0&1
\end{bmatrix}
\mathbf{x}(t_n)+\mathbf{w}(t_n);\; n=1,2\dots,N-1.
\end{equation}
The covariance matrix of $\mathbf{w}(t_n)=[w_1(t_n)\: w_2(t_n)]^T$  is assumed to be time invariant and can be represented as \cite{suvorova2018phase}
\begin{equation} \label{eq3}
    E[\mathbf{w}\mathbf{w}^T]= 
    \left[ \begin{array}{cc}
        T^3/3 & T^2/2  \\
        T^2/2 & T
    \end{array}
    \right] \gamma^2.\\
\end{equation}
The derivation of \eqref{eq3} is given in Appendix A. 

For the rest of the paper the state path is denoted by $\mathbf{x}=\{\mathbf{x}(t_n)\}_{n=1,\dots,N}$, the frequency path by $\mathbf{f}=\{f(t_n)\}_{n=1,\dots,N}$, and the phase path by ${\bm{\phi}}=\{\phi(t_n)\}_{n=1,\dots,N}$.
Given observations $\mathbf{y}$, our decision  of either $H_0$ or $H_1$ is based on the posterior distribution $\Pr(k|\mathbf{y})$.
\section{Hidden Markov Model (HMM)}
\label{sec:HMM}
Before deriving $\Pr(k|\mathbf{y})$, we give a brief review of the widely accepted  HMM-based Viterbi algorithm.
In this method,  the hidden state variable is the frequency,
discretized into frequency bins, while the observations are divided into  time blocks.
The state dynamics capture the frequency wandering between the blocks into the transition probability matrix. 
A good choice for the transition probability in this context is 
\begin{equation}
\Pr(F_k|F_{k-1})=1/3, \quad |F_k-F_{k-1}|\le \text{ frequency bin width},
\end{equation}
and zero elsewhere. Here $F_k$ and $F_{k-1}$ denote the bin-discretized frequencies at neighbouring time blocks. 
The emission probability matrix is constructed by computing the absolute value of 
the DFTs for each time block.
The  method relies on the assumption that the  frequency is contained in one frequency bin within each block and  jumps only occur between  blocks. Hence, the size of the block is determined by the  dynamics of the underlying wandering frequency, as explained in Section \ref{rationale HMM}.
The hidden states are then estimated using the Viterbi algorithm  and the detection statistics are determined by the "Viterbi score". A detailed analysis can be found in \cite{suvorova2018phase}.
Unlike the HMM-based technique, where detection follows the estimation of the hidden frequency path, in  our approach we form the detection statistic directly by computing  $\Pr(k|\mathbf{y})$ using Bayes formula. This analysis is done in the next section.  

\section{Detection statistics based on posterior distribution}
\label{sec:derive posterior}
In this section, we derive the expression of  the posterior distribution $\Pr(k|\mathbf{y})$ of the detection statistic.
In order to evaluate $\Pr(k|\mathbf{y})$, the term $\Pr(\tilde{a},\mathbf{x},k=1|\mathbf{y})$ has to be computed, which provides us additional information about  parameters other than $k$. In other words, the two distinct objectives, estimation and detection,  normally done sequentially (as in the HMM described in Section~\ref{sec:HMM}), are integrated naturally into one single term through this posterior distribution.
\subsection{Structure of the detection statistic $\Pr(k|\mathbf{y})$}
Using the law of total probability and Bayes' Rule,  we write
\begin{subequations}
\label{Pr(k|Y)}
\begin{align}
 \Pr(k=0|\mathbf{y})   &\propto\Pr(\mathbf{y}|k=0)\Pr(k=0)\label{pr(k=0|y)}\\
     \Pr(k=1|\mathbf{y})   &  =\int\limits_{D}\int\limits_{\mathbb{C}} \Pr(\tilde{a},\mathbf{x},k=1|\mathbf{y}) {d\tilde{a}d\mathbf{x}} \label{pr(k=1|y)},
     \end{align}
\end{subequations}
where 
 $\mathbb{C}$ denotes  the complex numbers and  $D$  denotes the domain for $\mathbf{x}$.
Marginalizing out the parameters in \eqref{pr(k=1|y)} is nontrivial since $\mathbf{x}$ is high dimensional. Hence, we have to evaluate the integrand $\Pr(\tilde{a},\mathbf{x},k=1|\mathbf{y})$ by computing the posterior estimate for the parameters in $H_1$ space.
\subsection{Prior distributions}
For later use, we specify the  prior distributions for all  parameters used in  the algorithm. The prior for $k$ is assumed to be Bernoulli distributed with a tunable parameter $1-\alpha$, that of  $\tilde{a}$ under $k=1$, i.e., when the signal exists,  is chosen to be a complex Gaussian distribution, with mean $0$ and variance $\Delta$, i.e., $\tilde{a} \sim \mathcal{CN}(0,\Delta)$. Usually we set $\Delta$ to be a large number compared to $\sigma^2$ to reflect our initial uncertainty.
As stated earlier $\phi(t_1)=0$  and  $f(t_1)$ is chosen to be uniformly distributed on the frequency interval  $(0,U)$, i.e., $U\sim \mathcal{U}(0,U)$ with $0< U\le 1/T$, where $U$ denotes the bandwidth
  of $\mathbf{y}_c$ and $1/T$ is the sampling rate. The bandwidth is either known, or, as  here, assumed to be equal to the  Nyquist frequency, so that $U=1/T$, although this is not very critical.\footnote{In gravitational wave applications, $U$ is usually much smaller than the sampling rate. A typical continuous wave search is conducted over sub-bands of $\sim 1\,{\rm Hz}$ to facilitate handling the large volume of data involved, compared to the sampling frequency $\gtrsim 1\,{\rm kHz}$. Continuous wave signals from neutron stars are expected to be quasimonochromatic, with intrinsic frequency bin width $\lesssim 10^{-6}\,{\rm Hz}$ \cite{Riles_2013}.}
\subsection{Main result}
The main result of this work is the following theorem.
\begin{theorem}
The posterior for $k$ is 
\begin{subequations}
\label{eq:all}
\begin{align}
\Pr(k=0|\mathbf{y}) &\propto \alpha W_0 \label{eq:11a}\\
\Pr(k=1|\mathbf{y}) &\propto (1-\alpha)W_0\int\limits_{D}\int\limits_{\mathbb{C}} W_f W_{\tilde{a}}d\tilde{a}d\mathbf{x}\label{eq:11b}\\
\Pr(\tilde{a},\mathbf{x},k=1|\mathbf{y}) &\propto  (1-\alpha)W_0W_fW_{\tilde{a}}, \label{eq:11c}
\end{align}
\end{subequations}
with
\begin{equation}
\label{eq:k0}
W_0=\frac{1}{(\pi \sigma^2)^{N}}\exp\Bigl(-\frac{1}{\sigma^2} \mathbf{y}^H\mathbf{y}\Bigr), 
\end{equation}
where the superscript $H$ denotes  conjugate transpose,
\begin{equation}
\label{Wf}
W_f=\Bigl(\frac{q\sigma^2}{U\Delta}\Bigr)\exp[\eta(\mathbf{x})],\;f(t_1)\in(0,U),
\end{equation}
and
\begin{equation}
\label{eq:wa}
W_{\tilde{a}}=\frac{1}{\pi\sigma^2 q} \exp\Bigl[-\frac{1}{\sigma^2q}|\tilde{a}-\bar{a}|^2\Bigr],
\end{equation}
where
\begin{subequations}
\begin{align}
    q& =(N+\sigma^2/\Delta)^{-1} \label{eq:q}\\
    \bar{a} &= q\mathbf{D}_{f}^H\mathbf{y} \label{eq:abar}\\
    \mathbf{D}_{f}&=\exp(j2\pi \bm{\phi})\\
    \eta(\mathbf{x}) &= \frac{q}{\sigma^2}(\mathbf{D}_{f}^H\mathbf{y})^H(\mathbf{D}_{f}^H\mathbf{y})=\frac{|\bar{a}|^2}{\sigma^2q}.\label{eta}
\end{align}
\end{subequations}
\end{theorem}

\begin{proof}
The likelihood $W_0$ is
\begin{equation}
W_0\overset{\triangle}{=}\Pr(\mathbf{y}|k=0)=\frac{1}{(\pi \sigma^2)^{N}}\exp\Bigl(-\frac{1}{\sigma^2} \mathbf{y}^H\mathbf{y}\Bigr).
\end{equation}
As we  show later, our algorithm does not require numerical computation of  $W_0$ because it cancels out.

We have
\begin{equation}
    \Pr(\tilde{a},\mathbf{x},k=1|\mathbf{y})\propto \Pr(\tilde{a},\mathbf{x},k=1)\Pr(\mathbf{y}|\tilde{a},\mathbf{x},k=1)\label{pr(a,x,k=1|y)},
\end{equation}
where the likelihood term $\Pr(\mathbf{y}|\tilde{a},\mathbf{x},k=1)$ is rewritten as
\begin{equation}
    \frac{1}{(\pi \sigma^2)^{N}} \exp\Bigl[-\frac{1}{\sigma^2} (\mathbf{y}-\tilde{a}\mathbf{D}_{f}) ^H(\mathbf{y}-\tilde{a}\mathbf{D}_{f})\Bigr],
\end{equation}
with 
\begin{equation}
\mathbf{D}_{f}=\exp(j2\pi \bm{\phi}).
\end{equation}
The other factor  $\Pr(\tilde{a},\mathbf{x},k=1)$ is further expanded to
\begin{equation}
    \Pr(\tilde{a},\mathbf{x},k=1)=\Pr(k=1)\Pr(\tilde{a},\mathbf{x}|k=1)\label{pr(a,x,k=1)},
\end{equation}
where $\Pr(\tilde{a},\mathbf{x}|k=1)$ denotes the prior distribution for $\tilde{a}$ and $\mathbf{x}$ under the model $k=1$. As for the prior distribution of $\mathbf{x}$, given an initial state of the path $\mathbf{x}(t_1)$, the statistical representation of the whole state path $\mathbf{x}$ is determined by the model according to \eqref{origin model}.
Based on the above analysis and the prior distributions, we rewrite  \eqref{pr(a,x,k=1|y)} as
\begin{equation}
\label{eq:k1}
\begin{split}
\Pr(\tilde{a},\mathbf{x},k=1|\mathbf{y})&\propto \frac{1-\alpha}{U  \pi \Delta (\pi \sigma^2)^{N}} \exp\left(-\tilde{a}^H \tilde{a}/\Delta\right)
\\
&\times\exp\Bigl[-\frac{1}{\sigma^2} (\mathbf{y}-\tilde{a}\mathbf{D}_{f}) ^H(\mathbf{y}-\tilde{a}\mathbf{D}_{f})\Bigr].
\end{split}
\end{equation}
To assist MCMC sampling we now  expand the quadratic form in \eqref{eq:k1} and after some algebraic  manipulations,  rewrite it in the following way
\begin{equation}
\label{eq:p}
\begin{aligned}
&\exp\Bigl[-\frac{1}{\sigma^2} (\mathbf{y}-\tilde{a}\mathbf{D}_{f}) ^H(\mathbf{y}-\tilde{a}\mathbf{D}_{f})\Bigr]
\exp\Bigl(-\tilde{a}^H\tilde{a}/\Delta\Bigr)\\
&=\exp\Bigl[-\frac{1}{\sigma^2 q}(\tilde{a}-\bar{a})^H (\tilde{a}-\bar{a}))
 \Bigr]\exp\Bigl(-\frac{1}{\sigma^2}\mathbf{y}^H\mathbf{y}\Bigr)\\
 &\times\exp\Bigl[\frac{q}{\sigma^2}(\mathbf{D}_{f}^H\mathbf{y})^H(\mathbf{D}_{f}^H\mathbf{y})\Bigr],\\
\end{aligned}
\end{equation}
with 
\begin{subequations}
\begin{align}
    q&=(\mathbf{D}_{f}^H \mathbf{D}_{f}+\sigma^2/\Delta)^{-1} =(N+\sigma^2/\Delta)^{-1} \\
    \bar{a} &= q\mathbf{D}_{f}^H\mathbf{y}. 
\end{align}
\end{subequations}
Notice that $\bar{a}$ is simply  the least squares solution of $\tilde{a}$ in \eqref{k=1data}
 for a given  $\mathbf{D}_{f}$. We now define
\begin{equation}
\label{wtildea}
W_{\tilde{a}}\overset{\triangle}{=}\Pr(\tilde{a}|\mathbf{D}_{f},\mathbf{y},k=1)=\frac{1}{\pi\sigma^2 q} \exp\Bigl[-\frac{1}{\sigma^2q}|\tilde{a}-\bar{a}|^2\Bigr]
\end{equation} 
 as the normal distribution of $\tilde{a}$ given a specific draw $\mathbf{D}_{f}$, with mean $\bar{a}$ and variance $\sigma^2q$. Equation~\eqref{eq:wa} reflects our uncertainty of $\tilde{a}$ relative to $\bar{a}$ in consequence of  the observation noise. 
Combining the remaining terms  of \eqref{eq:k1} and \eqref{eq:p}, we obtain
\begin{equation}
\label{eq:wf}
\begin{aligned}
W_f&\overset{\triangle}{=} \frac{q\sigma^2}{U \Delta}\exp\Bigl[\frac{q}{\sigma^2}(\mathbf{D}_{f}^H\mathbf{y})^H(\mathbf{D}_{f}^H\mathbf{y})\Bigr]\\&=\frac{ q\sigma^2}{U \Delta}\exp\Bigl(\frac{|\bar{a}|^2}{\sigma^2q}\Bigr) \\&=\frac{ q\sigma^2}{U \Delta}\exp\eta(\mathbf{x}).
\end{aligned}
\end{equation} 
Specifically, $\eta(\mathbf{x})$ 
can be interpreted as the signal-to-noise ratio (SNR) evaluated along a sampled state path $\mathbf{x}$. 
Combining ~\eqref{eq:k1}~\eqref{wtildea} and \eqref{eq:wf}, we obtain the formulae for $\Pr(k=0|\mathbf{y})$ and $\Pr(k=1|\mathbf{y})$ as given in \eqref{eq:11b} \eqref{eq:11c}.
\end{proof}
\begin{corollary}
The posterior distribution for $k$ is approximated by
\begin{equation}
\label{eq:prk}
\Pr(k|\mathbf{y}) = \left \{
\begin{array}{lr}
     \frac{\alpha}{\alpha + (1-\alpha)W}, \ k=0 \\
     \frac{(1-\alpha)W}{\alpha + (1-\alpha)W}, \ k=1
\end{array}
\right.
\end{equation}
with 
\begin{equation}
W\approx\Bigl(\frac{\sigma^2}{N\Delta}\Bigr)^{1/2} \exp\Big[\max
\limits_\mathbf{x}[\eta(\mathbf{x})]\Big].
\end{equation}
\end{corollary}
\begin{proof}
To further evaluate the integral term defined in \eqref{eq:11b} , we define
\begin{subequations}
\begin{align}
\label{eq:W}
W&\overset{\triangle}{=}\int\limits_{\mathbb{C}} W_{\tilde{a}}d\tilde{a}\int\limits_{D} W_f d\mathbf{x}\\
&\approx\Bigl(\frac{\sigma^2}{N\Delta}\Bigr)^{1/2} \exp\Big[\max
\limits_\mathbf{x}[\eta(\mathbf{x})]\Big].\label{28b}
\end{align}
\end{subequations}
The first integral in \eqref{eq:W} is equal to  unity. Evaluation of  the second integral
uses  the Laplace approximation,  based on the assumption  that the integrand is strongly and singly peaked.
The expression $\max\limits_\mathbf{x}[\eta(\mathbf{x})]$ in \eqref{28b} is  easy to evaluate for constant frequency signals through the Fourier transform $\mathcal{F}(\mathbf{y})$, i.e.,
\begin{equation}
\eta(\mathbf{x})=|\mathcal{F}(\mathbf{y})|^2/\sigma^2,
\end{equation}
but to compute \eqref{eq:W} for a wandering path, we resort to the MCMC algorithm in \ref{section: MCMC}.
\end{proof}
\section{MCMC algorithm}
\label{section: MCMC}
\subsection{Basic principle}
Formally, the posterior probability of a parameter $\mu$, given data $D$ is given by
\begin{equation}
\label{Bayes}
    \Pr(\mu|D)=\frac{\Pr(D|\mu)\Pr(\mu)}{\int_{\mu\in \mathbb{S}}\Pr(D|\mu)\Pr(\mu)d\mu}
\end{equation}
where $\mathbb{S}$ is the domain of the parameters and $\Pr(\mu)$ and $\Pr(D|\mu)$ are the prior probability and likelihood probability, respectively. Equation \eqref{Bayes} is often hard to compute analytically because of the potentially high dimensional integration appearing in the denominator.

MCMC provides a way to compute \eqref{Bayes} without evaluating the denominator by the construction of  a Markov chain with equilibrium distribution as in \eqref{Bayes}. After MCMC converges, samples drawn from the Markov chain can be treated as random samples drawn from the true posterior distribution. The Maximum a Posteriori (MAP) estimate or other statistical quantities  can then be approximated using these ensemble samples.

Specific algorithms designed for our scenario are described  in detail in the following sections.
\subsection{Sampling rules}
\begin{figure}[!h] 
    \centering
    \begin{tikzpicture}
\node[state] (q1) {$k^{i-1}=0$};
\node[state, right of=q1] (q2) {$k^{i-1}=1$};
\draw [-Latex]
(q1) edge[loop above]  node{$s^i=0$ / \say{stay at $H_0$}} (q1)
(q2) edge[loop above] node{$s^i=1$ / \say{update}} (q2)
(q1) edge[bend left, above] node{$s^i=1$ / \say{birth}} (q2)
(q2) edge[bend left, below] node{$s^i=0$ / \say{death}} (q1);
\end{tikzpicture} 
\caption{Diagram of the finite state machine to generate MCMC proposals} 
\label{fig:fsm}
\end{figure}
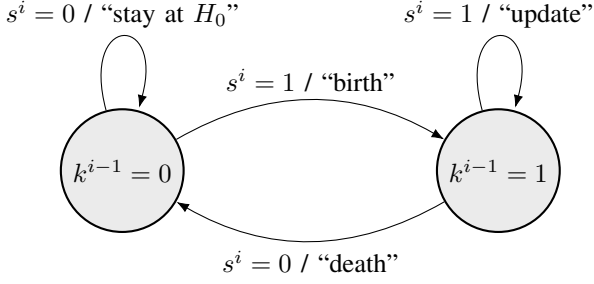

In this section, based on \eqref{eq:all}, we construct the specific MCMC algorithm for computing $\Pr(k|\mathbf{y})$. We use $(\cdot)^i$ to denote the value at the  $i$th iteration and $(\cdot)'$ to denote the proposed value. Firstly, we  specify a recipe for proposals to either switch between different hypotheses (``birth''/ ``death'') or explore parameter space (``update'') under hypothesis $H_1 (\text{or } k=1)$.
That is,  we build a finite state machine (Fig.~\eqref{fig:fsm}) of which the  output returns the instruction for the  next move. Specifically, we introduce a new random Boolean variable $s^i$. It evolves as a Markov chain with a given transition matrix $\mathbf{\Gamma}$, as
defined in Table \eqref{parametrs in algorithms}. The value of $s^i$ combined with the previous $k$ value $k^{i-1}$ gives  the  instruction to either  jump between $H_0$ and $H_1$ or search within $H_1$.
 Algorithms \ref{algorithm2}-\ref{algorithm7} in Appendix~B describe the implementation of MCMC as well as ``birth'', ``death'' and ``update'' in more detail.
From \eqref{eq:all}, the acceptance ratio
for  traversing between
``birth'', ``death'' and ``update'' proposals 
at the $i$th iteration are, respectively,
\begin{subequations}
\begin{align}
    r_\text{birth} &\overset{\triangle}{=} \frac{(1-\alpha)W_f^{'}W_{\tilde{a}}^{'}}{\alpha}\\
    r_\text{death} &\overset{\triangle}{=} \frac{\alpha}{(1-\alpha)W_f^{i-1}W_{\tilde{a}}^{i-1}}\\
    r_\text{update} &\overset{\triangle}{=}\frac{W_f^{'}W_{\tilde{a}}^{'}}{W_f^{i-1}W_{\tilde{a}}^{i-1}},
\end{align}
\end{subequations}
where $W_f^{'}$ and $W_{\tilde{a}}^{'}$ are evaluated at the proposed (perhaps rejected) sample values  ${\mathbf{x}}'$ and $\tilde{a}'$ at the current iteration,  according to \eqref{Wf} and \eqref{eq:wa},
and $W_f^{i-1}$ and $W_{\tilde{a}}^{i-1}$ are obtained from previous values $\mathbf{x}^{i-1}$ and $\tilde{a}^{i-1}$.
The acceptance probabilities are then
\begin{subequations}
\begin{align}
    A_\text{birth} &\overset{\triangle}{=} \min(1,r_\text{birth}) \label{abirth}\\
    A_\text{death} &\overset{\triangle}{=} \min(1,r_\text{death}) \label{adeath}\\
    A_\text{update} &\overset{\triangle}{=} \min(1,r_\text{update})\label{aupdate}.
\end{align}
\end{subequations}

\subsection{Knot-interpolation Scheme to  Reduce  Parameter Dimension}
High dimensionality of the parameter space may cause convergence problems for MCMC \cite{jones2001honest}. 
To alleviate this problem. for  both  the HMM and MCMC approaches, the time series $\mathbf{y}$ (of length $N$ and sampled at time intervals of length $T$)  is partitioned into consecutive blocks of  equal time duration,  $T_b$, though the chosen lengths of these blocks will differ between the two approaches. The endpoints of these blocks, at intervals of time length $T_b$ are called the \emph{knots}.  The number of  blocks is $N_b$, so that $NT=N_bT_b$. As described in Section~\ref{sec:HMM}, the  HMM-based method requires calculation of the DFT of each block to produce the emission probabilities, whereas in the MCMC approach a quadratic interpolation between the knots is used to approximate the time series and reduce dimensionality, as we only sample $\mathbf{x}$'s at the knots.
The  interpolation between the knots is performed in the following way
\begin{equation}
\label{interpolate}
\left\{
\begin{array}{ll}
\tilde{{\phi}}(t_{Mm+1+\ell})&=\phi(t_{Mm+1})+f(t_{Mm+1})T\ell\\
&+\frac{1}{2}b_1^m (T\ell)^2+\frac{1}{3}b_2^m(T\ell)^3\\
\tilde{f}(t_{Mm+1+\ell})&=f(t_{Mm+1})+b_1^mT\ell+b_2^m(T\ell)^2, 
\end{array}\right.
\end{equation}
where $M=N/N_b$   and $\ell = 0,\ldots,M-1$  denote time epochs within the $m$th block and $T_b=MT$ is the time duration of one block, as mentioned above.
Notice that the resulting interpolated path $\{\tilde{\phi}(t_n),\tilde{f}(t_n)\}_{n=1,\dots,N}$ is of length $N$.
The continuity of the interpolated path is ensured by
 solving~\eqref{interpolate} for  $b_1^m$ and $b_2^m$ using the values at the knots.
 
 Denoted by the function ``Interp", this  procedure is described in Appendix B Algorithm \ref{interp}.
The dynamics between the knots  is identical to  the dynamics in \eqref{origin model} with $T$ replaced by $T_b$.
The rationale for choosing $T_b$ (or equivalently, $N_b$) is discussed in Section~\ref{rationale MCMC}.

 From now  we  focus on generating the sequence of values, $\mathbf{x}_{N_b}$, of the path at the knots,   in \say{birth} and \say{update} scenarios. For simplicity of notation, we indicate the elements of $\mathbf{x}_{N_b}$ by $\mathbf{x}_{N_b}(j)$ for $j=1,\dots,N_b$.

\subsection{Generating a sample path $\mathbf{x}_{N_b}'$ (``birth'')}\label{birth}
For the MCMC \say{birth} procedure, we generate a random path $\mathbf{x}'_{N_b}$ with length $N_b$ and individual elements
\begin{equation} 
\label{Reducedpath}
\mathbf{x}'_{N_b}(j+1) = 
\left[ \begin{array}{cc}
     1&T_b  \\
     0&1
\end{array}
\right]
\mathbf{x}'_{N_b}(j)+\mathbf{w}(j).
\end{equation}
  The noise term $\mathbf w(j)$ is calculated in the same way as in \eqref{eq3}, with $T$ replaced by $T_b$. Here we define
\begin{equation}
\label{eq:covmatrix}
\mathbf{C}\overset{\triangle}{=}E[\mathbf{w}(j)\mathbf{w}(j)^T]= \gamma^2\left[ \begin{array}{cc}
     T_b^3/3&T_b^2/2  \\
     T_b^2/2&T_b
\end{array}\right],
\end{equation}
for $j=1,2\dots,N_b-1$.
\subsection{Updating a proposal path $\mathbf{x}_{N_b}'$ from previous path $\mathbf{x}_{N_b}^{i-1}$}\label{update}
It is  important in the MCMC algorithm to  formulate a good update proposal that specifies the probability of moving to a new point in parameter space --- a stochastic path  $\mathbf{x}_{N_b}'$, given previous location $\mathbf{x}_{N_b}^{i-1}$.
In this work,  we developed a unique approach to this problem, to be described here.
The desired new  path $\mathbf{x}_{N_b}'$ should possess the following properties: 
\begin{enumerate}[(i)]
\item  it should  obey the  state dynamics model  in \eqref{Reducedpath}.
\item  it should be ``close'' to the previous path to avoid a large chance of rejection. This is especially critical when the samples are near the peak of the posterior probability density function;
\item  the distance between paths should be controllable, to facilitate a flexible sampling scheme such as, for example,  to be able to increase the  convergence rate or to escape from  local extrema.
\end{enumerate}
Consequently, we want to control the Euclidean distance $||\mathbf{x}_{N_b}'-\mathbf{x}_{N_b}^{i-1}||$. To achieve this, 
 we expand \eqref{Reducedpath}  as
\begin{equation}
\label{eq:old path}
\begin{bmatrix}    
\mathbf{x}'_{N_b}(1)\\
\mathbf{x}'_{N_b}(2)\\
\vdots\\
\mathbf{x}'_{N_b}(N_b)
\end{bmatrix}=
\begin{bmatrix}
\mathbf{I}_2&0&\dots&0\\
\mathbf{F}&\mathbf{I}_2&\dots&0\\
\vdots&\vdots&\ddots&\vdots\\
\mathbf{F}^{N_b-1}&\mathbf{F}^{N_b-2}&\dots&\mathbf{I}_2
\end{bmatrix}
\begin{bmatrix}
\mathbf{x}'_{N_b}(1)\\
\mathbf{w}(1)\\
\vdots\\
\mathbf{w}(N_b-1)
\end{bmatrix}
\end{equation}
with $\mathbf{w}(j)\sim \mathcal{N}(\mathbf{0},\mathbf{C}),\ \mathbf{F}= \left[ \begin{array}{cc}
     1&T_b  \\
     0&1
\end{array}\right]$.
To generate a new path,  each $\mathbf{w}(j)$ is replaces by  a new random vector $\mathbf{L}\mathbf{q}(j)$, where  $\mathbf{L}\mathbf{L}^{T}=\mathbf{C}$ is the Cholesky decomposition, and  $\mathbf{q}(j)\sim \mathcal{N}(0,\mathbf{I}_2)$ is a bivariate normal vector with unit covariance matrix.
Now \eqref{eq:old path} for the new path becomes
\begin{subequations}
\label{M1M2}
\begin{align}
\begin{bmatrix}    
\mathbf{x}'_{N_b}(1)\\
\mathbf{x}'_{N_b}(2)\\
\vdots\\
\mathbf{x}'_{N_b}(N_b)
\end{bmatrix}&=
\begin{bmatrix}
\mathbf{I}_2&0&\dots&0\\
\mathbf{F}&\mathbf{I}_2&\dots&0\\
\vdots&\vdots&\ddots&\vdots\\
\mathbf{F}^{N_b-1}&\mathbf{F}^{N_b-2}&\dots&\mathbf{I}_2
\end{bmatrix}
\begin{bmatrix}
\mathbf{L}\mathbf{L}^{-1}\mathbf{x}'_{N_b}(1)\\
\mathbf{L}\mathbf{q}(1)\\
\dots\\
\mathbf{L}\mathbf{q}(N_b-1)
\end{bmatrix}\\
&=
\underbrace{
\begin{bmatrix}
\mathbf{L}&0&\dots&0\\
\mathbf{F}\mathbf{L}&\mathbf{L}&\dots&0\\
\vdots&\vdots&\ddots&\vdots\\
\mathbf{F}^{N_b-1}\mathbf{L}&\mathbf{F}^{N_b-2}\mathbf{L}&\dots&\mathbf{L}
\end{bmatrix}}_{\mathbf{M}_2}
\begin{bmatrix}
\mathbf{L}^{-1}\mathbf{x}'_{N_b}(1)\\
\mathbf{q}(1)\\
\vdots\\
\mathbf{q}(N_b-1)
\end{bmatrix}\\
&=\mathbf{M}_2\left(\begin{bmatrix} \label{Qnoise}
\mathbf{L}^{-1}\mathbf{x}'_{N_b}(1)\\
0\\
\vdots\\
0
\end{bmatrix}+\begin{bmatrix}
\mathbf{q}(0)\\
\mathbf{q}(1)\\
\vdots\\
\mathbf{q}({N_b-1})
\end{bmatrix}\right) \\
&= \underbrace{\begin{bmatrix} \label{M1,M2}
\mathbf{I}_2\\
\mathbf{F}\\
\vdots\\
\mathbf{F}^{N_b-1}
\end{bmatrix}}_{\mathbf{M}_1}\mathbf{x}'_{N_b}(1)+\mathbf{M}_2\begin{bmatrix}
\mathbf{q}(0)\\
\mathbf{q}(1)\\
\vdots\\
\mathbf{q}(N_b-1)
\end{bmatrix},
\end{align} 
\end{subequations}
with $\mathbf{q}(0)=[0,\;0]^T$ and $\mathbf{q}(j)\sim \mathcal{N}(\mathbf{0},\mathbf{I}_2)$ for $j=1,\ldots,N_b-1$.
From \eqref{M1,M2}, we observe  that a stochastic state path $\mathbf{x}'_{N_b}$ depends purely on the random starting point $\mathbf{x}'_{N_b}(1)$ and  the random noise sequence $\mathbf{q}=[\mathbf{q}^T(0),\ldots, \mathbf{q}^T(N_b-1)]^T$ since matrices $\mathbf{M}_1$ and $\mathbf{M}_2$ are deterministic.
In our approach we  keep $\mathbf{x}_{N_b}'(1)=\mathbf{x}_{N_b}^{i-1}(1)$ and only perturb the random noise sequence $\mathbf{q}$. Under the requirement (i) above,  the mean and variance of the perturbed noise sequence need to be retained. Specifically, the steps for perturbing the noise sequence at the $i$th iteration are: given a previous path $\mathbf{x}_{N_b}^{i-1}$, first extract the  random part: $\mathbf{q}^{i-1}=\mathbf{x}^{i-1}_{N_b}-\mathbf{M}_1\mathbf{x}_{N_b}^{i-1}(1)$; then  generate a white noise perturbation sequence $\mathbf{q}'=[\mathbf{q}^T(0)\dots,  \mathbf{q}^T(N_b-1)]$, with $\mathbf{q}(0)=[0,\;0]^T$ and random vector $\mathbf{q}(j)$  with  zero mean and unit covariance $\text{cov}(\mathbf{q}(j))=\mathbf{I}_2$ for $j=1,\ldots, N_b-1$. The perturbation sequence $\mathbf{q}'$  is independent of $\mathbf{q}^{{i-1}}$, that is $E[\mathbf{q}' (\mathbf{q}^{{i-1}})^T]=\mathbf{0}$. We introduce a parameter $\beta$  and compute the new noise sequence as $\mathbf{q}^{i}=\mathbf{q}^{i-1}\cos{\beta}+\mathbf{q}'\sin{\beta}$. This perturbation scheme ensures that the new noise sequence has the required  mean and  variance, because of the identity $\cos^2\beta +\sin^2\beta =1$.  It is also apparent that $\cos\beta$ is the  correlation coefficient between each old $\mathbf{q}^{i-1}(j)$ and new $\mathbf{q}^i(j)$ for $j=1,\ldots N_b-1$, thus  the  tunable parameter $\beta$ helps control the "closeness" between the old and the new sequence, i.e.,  for smaller $\beta$, the correlation is greater, hence the  perturbation is smaller.

 This scheme has one problem related to 
  the lower triangular shape of the matrix $\mathbf{M}_2$.
The noise  in the new path $\mathbf{M}_2\mathbf{q}^i$ tends to accumulate along the path, meaning that   $||\text{cov}(\mathbf{q}^i(n))||>||\text{cov}(\mathbf{q}^i(m))||$ for $n>m$ as the iteration number $i$ increases.
Our \emph{ad hoc} solution  to this problem is the following. Instead of setting $\mathbf{x}_{N_b}'(1)=\mathbf{x}_{N_b}^{i-1}(1)$,
we start from a random  position $l\in\{1,\dots, N_b\}$  and let $\mathbf{x}_{N_b}'(l)=\mathbf{x}_{N_b}^{i-1}(l)$;  the sequence is then split into two, with one part propagating backward all the way to $\mathbf{x}'_{N_b}(1)$ and the other part propagating forward  until $\mathbf{x}'_{N_b}(N_b)$. This is achieved by replacing matrices $\mathbf{M}_1$ and $\mathbf{M}_2$ by  $\mathbf{M}_1'$ and $\mathbf{M}_2'$ as follows:
\begin{equation}
\label{M_1'}
\begin{aligned}
    \mathbf{M}_1'&=
    \begin{bmatrix}
    \mathbf{F}^{-(l-1)} &
    \mathbf{F}^{-(l-2)} &
    \dots &
    \mathbf{I}_2 &
    \mathbf{F} &
    \dots &
    \mathbf{F}^{N_b-l}
    \end{bmatrix}^T\\
    \mathbf{M}_2'&=\\
    &=
    \begin{bmatrix}
    \mathbf{L}&\mathbf{F}^{-1}\mathbf{L}&\dots& \mathbf{F}^{-(l-1)}\mathbf{L}&0&\dots&0\\
    0&\mathbf{L}&\dots& \mathbf{F}^{-(l-2)}\mathbf{L}&0&\dots&0\\
    \vdots&\ddots&\vdots&\vdots&\vdots&\ddots&\vdots\\
    0&\dots&\mathbf{L}&\mathbf{F}^{-1}\mathbf{L}&0&\dots&0\\
    0&\dots&0&\mathbf{L}&0&\dots&0\\
    0&\dots&0&\mathbf{F}L&\mathbf{L}&\dots&0\\
    \vdots&\ddots&\vdots&\vdots&\vdots&\ddots&\vdots\\
    0&\dots&0&\mathbf{F}^{N_b-l}\mathbf{L}&\mathbf{F}^{N_b-l-1}\mathbf{L}&\dots&\mathbf{L}
    \end{bmatrix}.
    \end{aligned}
\end{equation}
Notice that when $l=1$, we recover $\mathbf{M}_1'=\mathbf{M}_1$ and $\mathbf{M}_2'=\mathbf{M}_2$.
This still causes noise accumulation in $\mathbf{x}_{N_b}'$ for  elements away from $l$ in both directions, but  the random choice  of $l$ at each iteration mitigates the effect in the long run.

The correlation between previous and  proposed paths is
\begin{equation}
\label{correlation}
\begin{aligned}
    \mathbf{x}_{N_b}^{i-1} &=\mathbf{M}_1'\mathbf{x}_{N_b}^{i-1}(l)+\mathbf{M}_2'\mathbf{q}^{i-1}; \\
    \mathbf{x}_{N_b}' &=\mathbf{M}_1'\mathbf{x}_{N_b}^{i-1}(l)+\mathbf{M}_2'\mathbf{q}^{i-1}\cos{\beta}+\mathbf{M}_2' \mathbf{q}' \sin{\beta};\\
    \text{cov}(\mathbf{x}_{N_b}^{i-1},\mathbf{x}_{N_b}')&=E[\mathbf{M}_2'\mathbf{q}^{i-1}(\mathbf{M}_2'\mathbf{q}^{i-1}\cos{\beta}+\mathbf{M}_2' \mathbf{q}' \sin{\beta})^T]\\
    &=(\mathbf{M}_2')^2\cos{\beta}E[\mathbf{q}^{i-1}(\mathbf{q}^{i-1})^T]\\
    &=(\mathbf{M}_2')^2\cos{\beta}\mathbf{I}_{2\times N_b},
    \end{aligned}
\end{equation}
where we use $E[\mathbf{q}^{i-1}(\mathbf{q}^{i-1})^T]=\mathbf{I}_{2\times N_b}$, $E[\mathbf{q}^{i-1}\mathbf{q}'^T]=0$ and $E(\mathbf{x}_{N_b}^{i-1})=E(\mathbf{x}_{N_b}')=\mathbf{M}_1'\mathbf{x}_{N_b}^{i-1}(l)$. Equation \eqref{correlation} indicates how the correlation of previous and proposed paths can be tuned by  $\beta$.

 For completeness, the distance between neighbouring paths in the  $\mathcal{L}_2$ norm is bounded by
\begin{equation}
\begin{aligned}
    ||\mathbf{x}_{N_b}^{i-1}-\mathbf{x}_{N_b}'||&=||\mathbf{M}_2'\mathbf{q}^{i-1}(\cos \beta-1)+\mathbf{M}_2' \mathbf{q}' \sin \beta|| \\
    &\le 
   \sigma_{\mathbf{M}_2'} \sqrt{2N_b}(\cos\beta+\sin\beta-1),
\end{aligned}
\end{equation}
where $\sigma_{\mathbf{M}_2'}$ is the  largest singular value  of $\mathbf{M}_2'$.
The pseudocode of the method is provided in Appendix B, Algorithm \ref{algorithm7}.

\section{Numerical validation}
\label{sec:experiment}
\subsection{Description of synthetic data}
To test our MCMC algorithm,   a synthetic data sequence with length $N$ is generated  according to \eqref{k=1data}, \eqref{origin model} and \eqref{eq3}.
Parameters for synthetic data are given in Table~\eqref{synthetic data}. The starting frequency  $f(t_1)$ is chosen randomly from $(0,U)$ with $U=1$.  The true path  sequence $\mathbf{x}_\text{syn}~=~\{\mathbf{x}_{\rm syn}(t_n)\}_{n=1,\dots,N}$, with $\mathbf{x}_{\rm syn}(t_n)=[\phi_{\rm syn}(t_n),f_{\rm syn}(t_n) ]^T$, is randomly synthesised  according to the dynamics given in \eqref{origin model} and \eqref{eq3}; the complex-valued amplitude $\tilde{a}=|\tilde{a}| \exp(j\psi_0)$ is also chosen randomly from the distribution $\mathcal{CN}(0,\Delta)$, with $\psi_0 \sim \mathcal{U}(0,2\pi)$. The signal-to-noise ratio  is
$\text{SNR}=\frac{|\tilde{a}|}{\sigma}$. This SNR differs from the SNR  along a path defined in \eqref{eta}: $\eta(\mathbf{x})$.
\begin{table}
    \caption{Parameters for generating synthetic test data}
\centering
\setlength{\tabcolsep}{3pt}
    \begin{tabular}{ |m{0.8cm}|m{0.6cm}|m{0.95cm}|m{1.45cm}|m{1.8cm}|m{2cm}|m{2.2cm}|m{2.2cm}|}
    \hline
    &Data length & Sampling interval  &  Diffusion constant of Wiener process & Signal-to-noise-ratio & \makecell{standard deviation\\ of \\observation noise}\\
    \hline
        Symbol&$N$  &\makecell{$T=\frac{1}{U}$ \\(sec)} &\makecell{$\gamma_{\rm syn}$\footnotemark \\(Hz~sec$^{-1/2}$)}& \makecell{SNR} &\makecell{$\sigma$\\ (arbitrary units)} \\
\hline 
        Value&1000 & 1 & $1\times 10^{-4}$ & $\{0.1,0.15,0.2\}$ & $\sqrt{20}$\\
        \hline
    \end{tabular}
    \label{synthetic data}
\end{table}
\footnotetext{We differentiate $\gamma_{\rm syn}$ from $\gamma$, the first one is used in generating synthetic data, and the second one is the parameter of MCMC algorithms. The chosen value of $\gamma_{\rm syn}$ is to reflect the degree of randomness of the wandering frequency.}
\subsection{Description of MCMC  Parameters}
Prior distributions of unknown parameters and specific values are given in Table~\eqref{prior distribution} and Table~\eqref{prior distribution number} respectively. The number of blocks, $N_b$, is chosen from the set \{5,  20, 200, 500, 1000\} to investigate how it affects the runtime and detection performance. 
$T_b$ is determined  according to $T_bN_b=TN$.
The  parameters used in the implementation of the algorithms  are given  in Table~\eqref{parametrs in algorithms}. The factor $\beta=0.1$ is chosen experimentally to ensure a reasonable  MCMC acceptance rate.
\begin{table}
    \caption{Prior distribution of $\mathbf{x}_1,\ \tilde{a}, \ k,\ \mathbf{W}_j$}
\centering
\setlength{\tabcolsep}{3pt}
    \begin{tabular}{ |p{1.3cm}|p{1.1cm}|p{1.15cm}|p{1.1cm}|p{1.1cm}| p{1.0cm}| }
    \hline
        $\Pr(\phi(t_1)=0)$ & $\Pr(f(t_1))$ &$\Pr(\tilde{a})$&$\Pr(k=0)$&$\Pr(k=1)$&$\Pr(\mathbf{w}(j))$\\
\hline 
       $ 1$ & $\mathcal{U}(0,U)$ & $\mathcal{CN}(0,\Delta)$& $\alpha$ & $1-\alpha $ & $\mathcal{N}(\mathbf{0},\mathbf{C})$\\
       \hline
    \end{tabular}
    \label{prior distribution}
\end{table}
\begin{table}[h!]
    \caption{Values of prior distribution parameters}
    \setlength{\tabcolsep}{3pt}
\centering
    \begin{tabular}{ |p{1cm}|p{1.5cm}|p{0.4cm}|p{3cm}| }
    \hline
        $U$ (Hz) & $\Delta$ (arb.units) &$\alpha$&$\mathbf{C}$\\
\hline 
       $ 1$ & $1\times 10^2$ & $0.5$& \eqref{eq:covmatrix} with $\gamma =  10^{-4}$ (or $\gamma =  10^{-5}$) Hz~sec$^{-1/2}$\\
\hline
    \end{tabular}
    \label{prior distribution number}
\end{table}

\begin{table}[h!]
\caption{Parameters appearing in Algorithm 2-7}
\setlength{\tabcolsep}{3pt}
\centering
    \begin{tabular}{ |m{2cm}|m{2cm}|m{1.5cm}|m{2cm}|}
    \hline
    Number of iterations $N_{\text{iteration}}$ & Transition matrix $\mathbf{\Gamma}$ for $s^i$ & Correlation factor $\beta$ &Diffusion constant of Wiener process $\gamma$ \\
    \hline
        $1\times 10^5$ & $\begin{bmatrix}
0.5 & 0.5\\
0.5 & 0.5
\end{bmatrix}$ &$0.1$&$\{10^{-4},10^{-5}\}$\\
\hline 
\end{tabular}
    \label{parametrs in algorithms}
\end{table}
\subsection{MCMC-posterior distributions}
In Section \ref{sec:derive posterior} and \ref{section: MCMC}, we show that,  to compute the posterior $\Pr(k|\mathbf{y})$, we have to sample from $\Pr(k=1,\tilde{a},\mathbf{x}|\mathbf{y})$ as well as $\Pr(k=0|\mathbf{y})$. Hence, as a part of the detection algorithm, we  approximate the  MCMC-posterior for the state path  $\mathbf{x}$ as well, which is  achieved by simply collecting all of the sampled state paths under $k=1$. The  MCMC-MAP estimate is achieved by calculating the mode of these paths.  Since the HMM  algorithm  also estimates  $\mathbf{x}$, it is of interest to compare these two algorithms in terms of the estimated paths.


The MCMC-posterior for $k$: $\Pr(k|\mathbf{y})$ is approximated 
 by counting the number of occurrences of $k=1$ and $k=0$ respectively. A Neyman-Pearson type detector is constructed by comparing $\Pr(k|\mathbf{y})$ with a pre-defined threshold to determine detection.
 
 In the following sections, we compute the  MCMC-posterior distributions for $\mathbf{x}$ and $k$, respectively. Related performance criteria like estimation error and ROC curves are also presented.
\subsection{ Rationale for choosing the number of blocks (knots) }\label{rationale}
In this section we discuss    the reasoning behind the differences in the selection of  the number of blocks for  the HMM and MCMC
methods.

\subsubsection{For HMM}\label{rationale HMM}
In HMM, within one block, we perform an $M$-point DFT,  resulting in  frequency bins of width $\Delta f=U/M$, where $U$ and $M$, as in MCMC, denote the bandwidth of the signal and block size, respectively. The number of blocks $N_b$, or equivalently, $M$ is chosen such that
\begin{equation}
\label{freq bin}
    \Pr(\int_0^{T_b}df(s)\geq\Delta f)<\kappa,
\end{equation}
where $T_b=TM$ is the time duration within one block and $\kappa$ is restricted to be a small number.
With $f(t)$ undergoing the dynamics in ~\eqref{continuous}, the integral in~\eqref{freq bin} is
\begin{equation}
\label{freq wandering}
    \int_0^{T_b} df(s)=\gamma\Big(B(T_b)-B(0)\Big)\sim \mathcal{N}(0,\gamma^2 T_b),
\end{equation}
where $B(t)$ denotes the Wiener process at time $t$. 

We set the frequency bin width to be twice the standard deviation in \eqref{freq wandering}, i.e., $\Delta f = 2\gamma\sqrt{T_b}$. Combining the relation that $\Delta f=U/M$ and setting $\gamma=1\times 10^{-4}$ (Table.~\eqref{synthetic data}),  we finally choose $N_b=N/M=5$ for the HMM in the following experiments.
\subsubsection{For MCMC}\label{rationale MCMC}
To implement our MCMC algorithm, $N_b$ needs to be chosen beforehand. The optimum $N_b$, could be  computed by maximizing the likelihood ratio or deflection ratio, as in~\cite{veeravalli1991quadratic}.
In this section we describe an alternative, intuitive reasoning behind our choice of $N_b$ for the MCMC.

\begin{figure}
\begin{minipage}{.45\textwidth}
    \centering
    \includegraphics[width=1\linewidth]{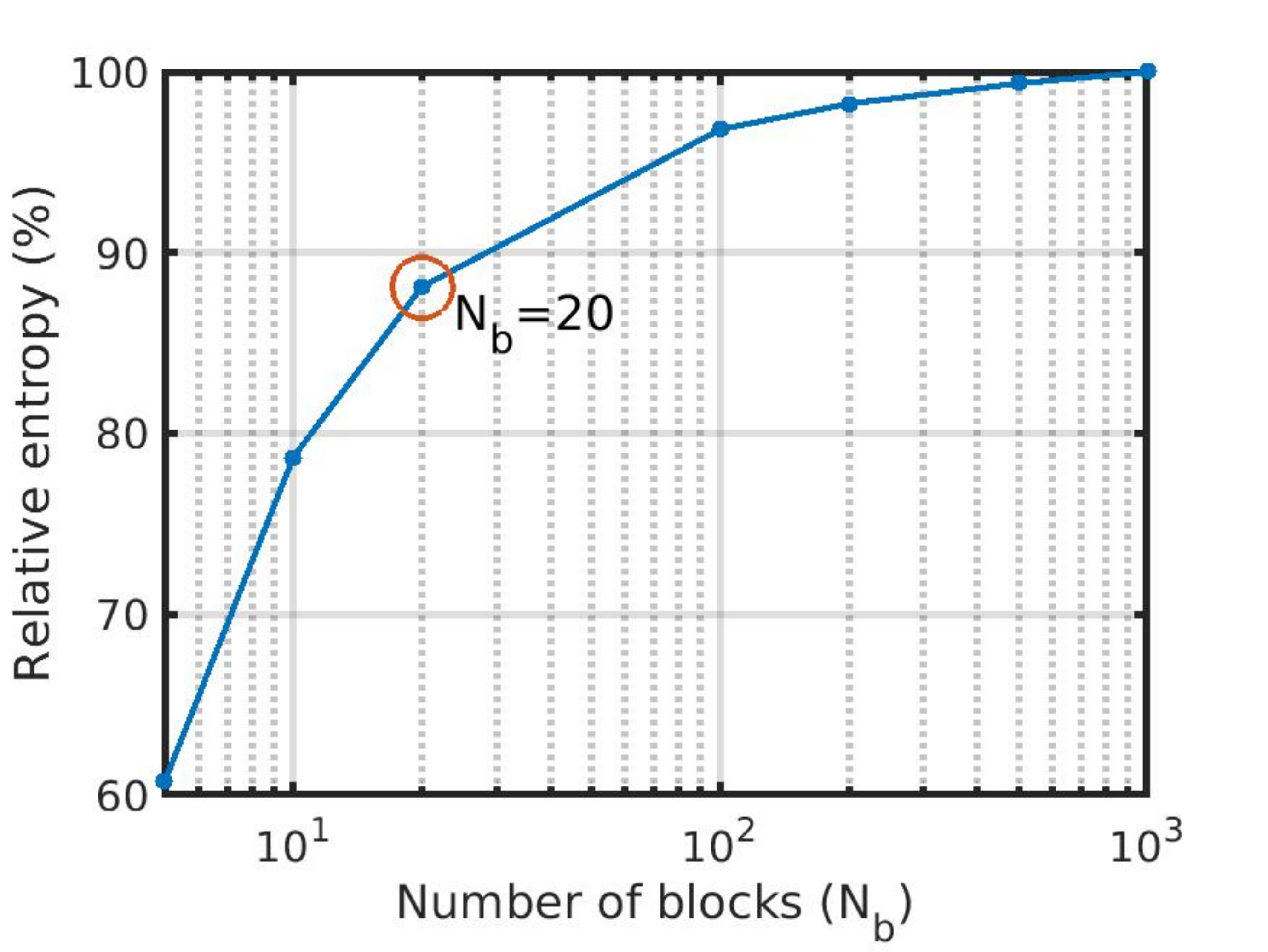}
\end{minipage}
\begin{minipage}{.45\textwidth}
    \centering
          \includegraphics[width=0.9\linewidth]{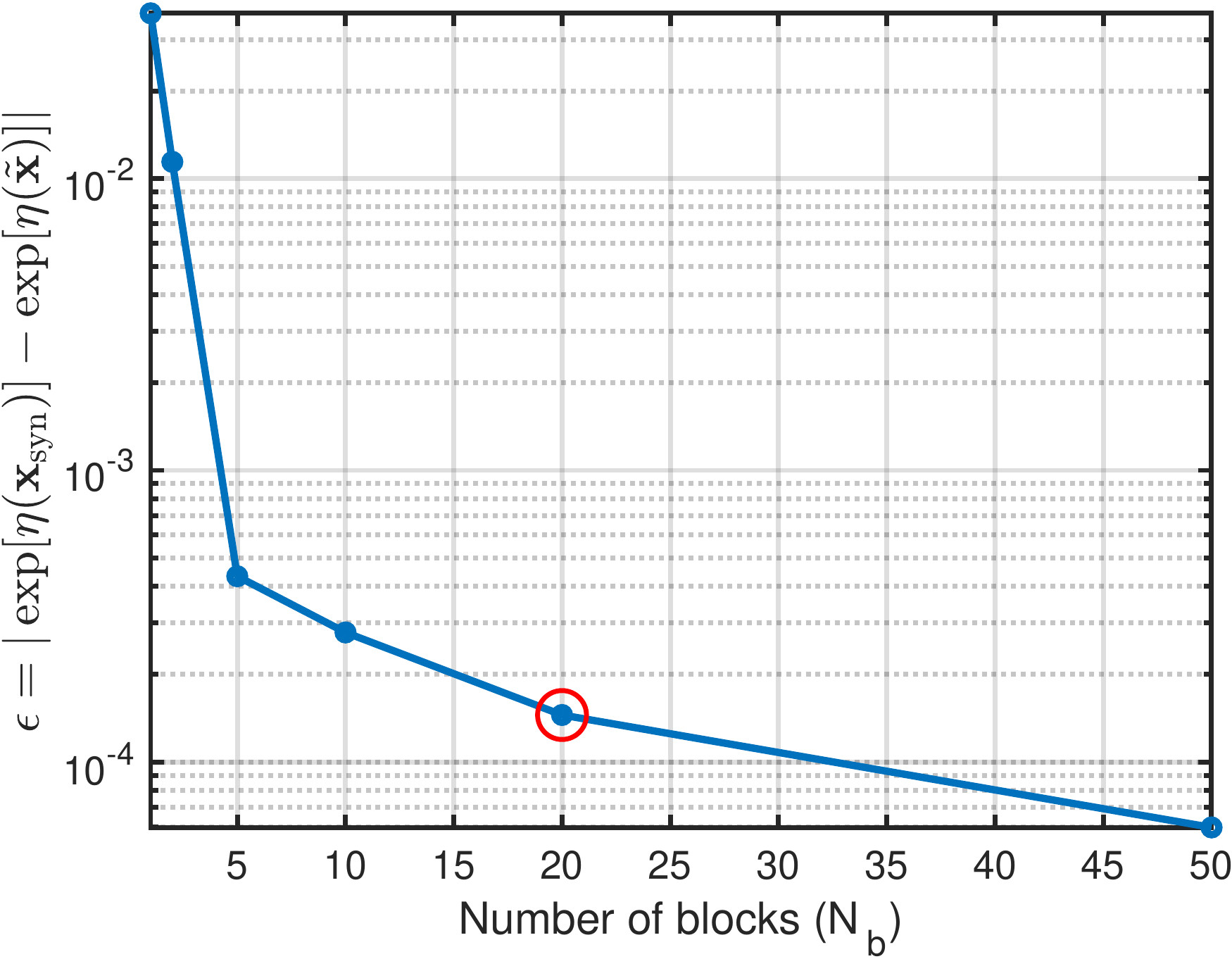}
\end{minipage}

\caption{Upper panel: Relative entropy between full dynamic matrix $\mathbf{M}_2^H$ and reduced matrices $\mathbf{M}_2$ as a function of $N_b$. The entropy associated with $N_b=20$ is circled in red, indicating $10$ percent information reduction. 
Lower panel: Absolute value of error caused by block approximation, i.e. coarse-graining error of the state path $\mathbf{x}$ versus block size, where the error is  $\epsilon=|\exp[\eta(\mathbf{x}_{\rm syn})]-\exp[\eta(\mathbf{\tilde{x}})]|$.}
\label{fig:PCA}
\end{figure}

Consider the  two  matrices  defined in \eqref{M1M2},
one of which is the full $2N\times 2N$ matrix $\mathbf{M}_2$ and the other is the $2N_b\times 2N_b$ matrix $\mathbf{M}_2^{b}$.
By design, $\mathbf{M}_2^b$ is constructed from $\mathbf{M}_2$ by keeping the rows and columns at the knots and removing the rest.
The difference in   ``information'' between  these two matrices is captured by the difference between the information theoretic ``Shannon entropy'' $H(N)$ and $H(N_b)$  of the singular values of $\mathbf{M}_2$ and $\mathbf{M}_2^{b}$ respectively. For any $\mathbf{M}_2^b$ we write
\begin{equation}
\label{eq:entrophy}
H(N_b)=-\sum\limits_i \xi_i^b\log(\xi_i^b),
\end{equation}  
where $\xi_i^b=\sigma_i/\sum\limits_{j=1}^{2N_b} \sigma_j^b$ and $\sigma_j^b$ is the $j$th largest non-zero singular value of $\mathbf{M}_2^b$. The entropy $H(N)$ is computed for the full matrix $\mathbf{M}_2$.  This entropy is effectively the same as the  von~Neumann entropy~\cite{simmons2017symmetric} in the context of symmetric matrices.
Accordingly, for a chosen set of parameters,  we compute the  entropy for a given  $N_b$  relative to $H(N)$ of the full matrix. The result is shown on the left hand panel in Fig.~\eqref{fig:PCA}.

To strengthen the entropy claim, we also directly evaluate $\exp[\eta(\mathbf{x}_{\rm syn})]$ (\eqref{eta}).
This is the  dominant term contributing to the Bayesian evidence (\eqref{eq:all}-\eqref{Wf}). Undoubtedly, if MCMC converges, there will be a high density of MCMC samples near the posterior probability, i.e., regions with larger   $\exp[\eta(\mathbf{x}_{\rm syn})]$. By probing how   $\exp[\eta(\mathbf{x}_{\rm syn})]$ varies as $N_b$ changes, we have a better understanding of the effect of the choice of $N_b$  on  the accuracy of the  posterior distribution estimation. In particular, while  decreasing  $N_{b}$, we measure the difference introduced by using   $\exp[\eta(\mathbf{\tilde{x}})]$ compared to $\exp[\eta(\mathbf{x}_{\rm syn})]$, where $\mathbf{\tilde{x}}$ again denotes the interpolated path from $N_b$ knots extracted from $\mathbf{x}_{\rm syn}$.
We first calculate the true value
\begin{equation}
\exp[\eta(\mathbf{x}_{\rm syn})]\overset{\triangle}{=}\exp\Bigl[\frac{q}{\sigma^2}(\mathbf{D}_{\rm syn}^H\mathbf{y})^H(\mathbf{D}_{\rm syn}^H\mathbf{y})\Bigr]
\end{equation}
 using noisy synthetic data $\mathbf{y}$ and $\mathbf{D}_{\rm syn}\overset{\triangle}{=}\exp(j2\pi {\bm{\phi}}_{\rm syn})$, where $\bm{\phi}_{\rm syn}$ stands for the synthetic phase path. This value provides an  upper bound. Then we compute
 \begin{equation}
\exp[\eta(\mathbf{\tilde{x}})]\overset{\triangle}{=}\exp\Bigl[\frac{q}{\sigma^2}(\tilde{\mathbf{D}}^H\mathbf{y})^H(\tilde{\mathbf{D}}^H\mathbf{y})\Bigr],
 \end{equation}
 where  $\tilde{\mathbf{D}}= \exp(j2\pi \tilde{\bm{\phi}})$. Here $\tilde{\bm{\phi}}$ is related to $\bm{\phi}_{\rm syn}$ by
 \begin{equation}
     \tilde{\bm{\phi}}(t_n)=\Big\{
     \begin{array}{lr}
      \bm{\phi}_{\rm syn}(t_n),\; (n-1)/M\in \mathbb{Z},    &  \\
      \text{Interp} \Big(\bm{\phi}_{\rm syn}(t_{M\lfloor\frac{n-1}{M}\rfloor+1}),\bm{\phi}_{\rm syn}(t_{M\lceil\frac{n-1}{M}\rceil+1})\Big),\; \text{else}  & 
     \end{array}
 \end{equation}
 for $n=1,\dots,N$ with $M=N/N_b$.
 The coarse-grained absolute value of the error in  calculating Bayesian evidence as a result of  interpolation is reflected in $\epsilon=|\exp[\eta(\mathbf{x}_{\rm syn})]-\exp[\eta(\mathbf{\tilde{x}})]|$. The error $\epsilon$ versus $N_b$ is plotted in the lower panel of Fig.~\eqref{fig:PCA}.
 Observe that for e.g., $N_b=20$ (red circle),  almost $90\%$  of the information is retained in the reduced $\mathbf{M}_2^b$ with $\epsilon< 10^{-3}$.
%
We believe that  computing the  entropy in \eqref{eq:entrophy} provides us an alternative way to select $N_b$. However, further investigation is required to justify the claim. In the following sections, we show in simulations that the choices of, for example, $N_b=20$, 
maintains MCMC performance in both estimation and detection, while saving computational resources significantly. This is reflected in Table~\eqref{tab:runtime}, where MCMC runtime averages over $10^3$ experiments  for different $N_b$'s with different number of iterations are reported, specifically for $N_\text{iteration}=5\times 10^3,\;10^4,\;5\times 10^4,\;10^5.$  The runtimes are computed on a 2.4GHz central processing unit (CPU).
 \begin{table}
\caption{\label{tab:runtime}Runtime as a function of number of blocks and number of iterations}
\setlength{\tabcolsep}{3pt}
\begin{center}
\begin{tabular}{|c|c|c|c|c|}
\hline
\diagbox{$N_b$}{Iterations} & $5\times 10^3$ & $10^4$ & $5\times 10^4$ & $10^5$ \\
\hline
20 & 0.2771s & 0.5717s & 2.9692s & 5.7394s \\
\hline
200 & 0.6078s &	1.1936s &	5.9269s &	12.8379s \\   
\hline
1000 & 2.4641s &	4.9219s &	25.4540s &	51.1769s \\
\hline
\end{tabular}
\end{center}
\end{table}

 \subsection{Estimation performance}
 Throughout this and the next section, we fix $N_b=5$ for the HMM (explained in Section \ref{rationale HMM}),
 and vary $N_b$ for the MCMC.
 \subsubsection{ MCMC-posterior for the state path: $\Pr(\mathbf{x}|\mathbf{y})$}
 
\begin{figure}[h!]
    \includegraphics[width=1\linewidth]{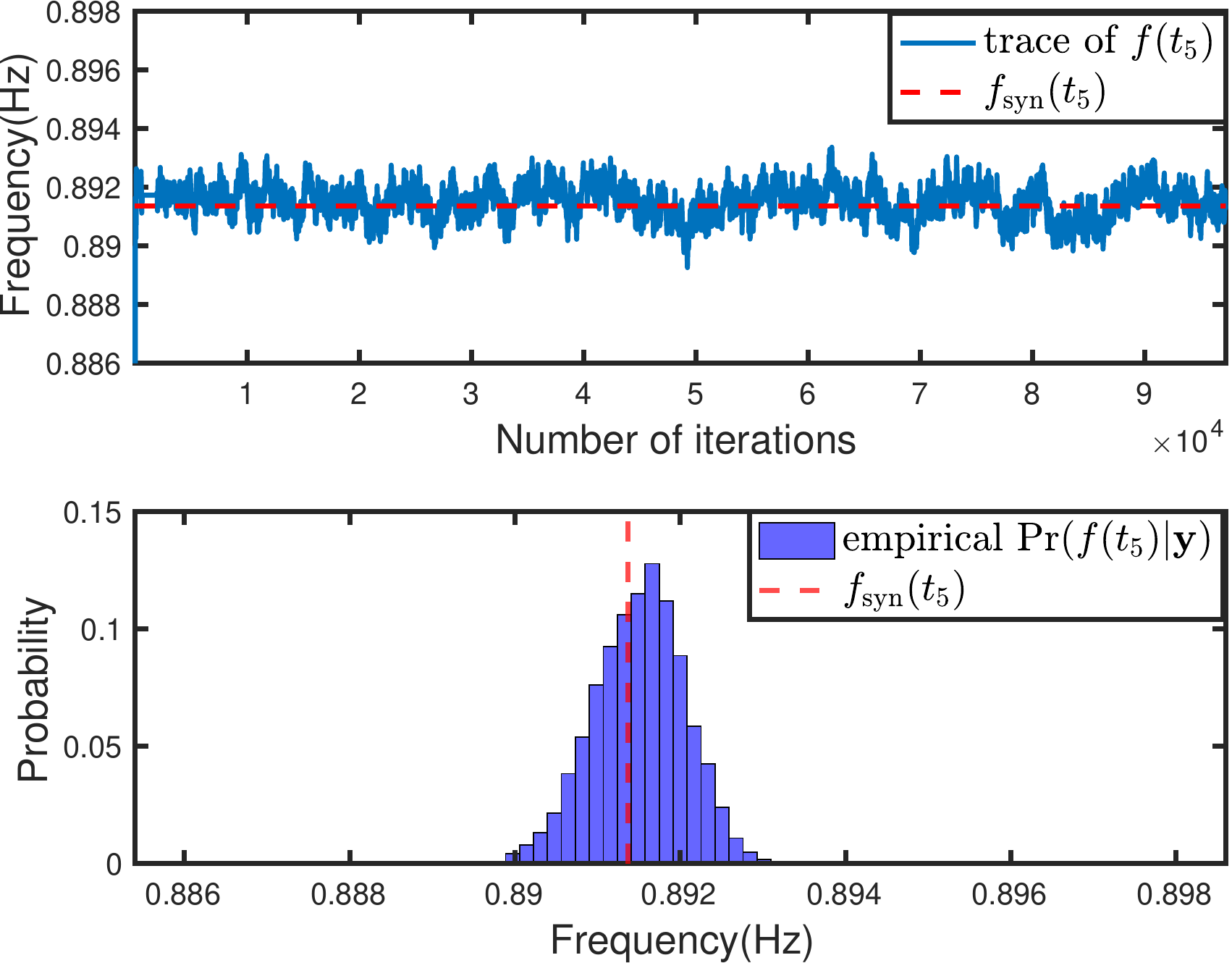}
    \includegraphics[width=1\linewidth]{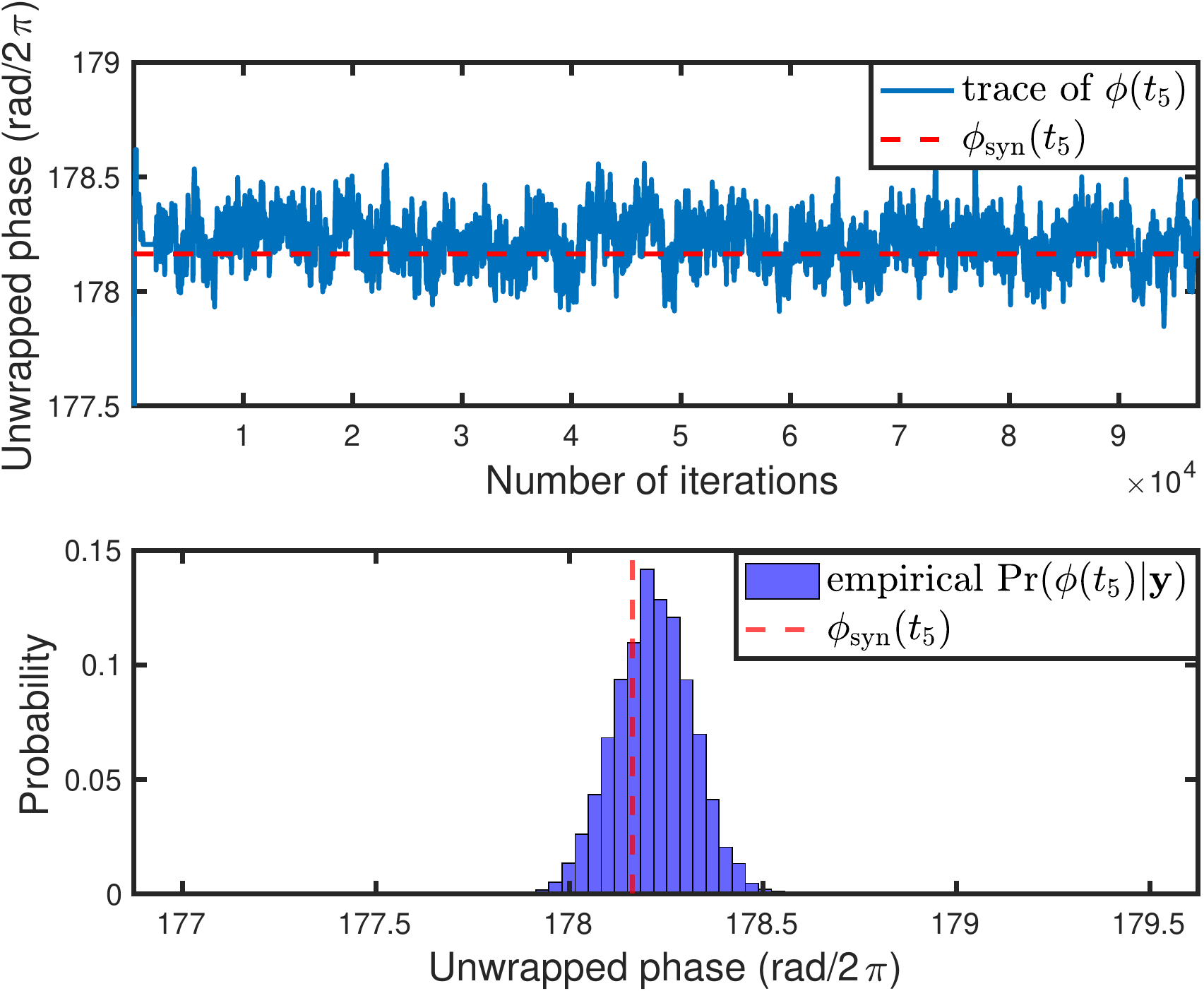}
 \caption{MCMC convergence for $k=1$ and SNR $=0.15$. First panel: Trace plot of $f(t_5)$. Second panel:  MCMC-posterior distribution $\Pr(f(t_5)|\mathbf{y})$. Third panel: Trace plot of $\phi(t_5)$. Fourth panel:  MCMC-osterior distribution $\Pr(\phi(t_5)|\mathbf{y}),\text{where }\mathbf{y} \text{ is generated from model } k=1  $. Injected parameters of synthetic data: $f_{\rm syn}(t_5)=0.8914$Hz, $\phi_{\rm syn}(t_5)=178.2324$ rad/$2\pi$, SNR~=~0.15. }
 \label{fig:Pr(x|Y)}
\end{figure}

 In Fig.~\eqref{fig:Pr(x|Y)},  a cross-section of the  MCMC-posterior $\Pr(\mathbf{x}|\mathbf{y})$ at time instant $t_5$  for SNR $= 0.15$ and $N_b=20$ is shown. The performance  at other epochs is similar.

Trace plots and histograms for $f(t_5)$ and $\phi(t_5)$, respectively, are shown. By definition, trace plots show the sampled values of a parameter over time. They reflect whether and how fast MCMC converges  in distribution.
Starting from a random initial point, MCMC converges after about $10^3$ iterations.
This, so called  ``burn in'' period is seen in the top and third panels in Fig.~\eqref{fig:Pr(x|Y)},  compressed into the left edge of the plots.
 After the  ``burn in'' period, the samples drawn from the MCMC have values centered around the true value, with bias less than $0.0005$ Hz ($0.05$ percent of the bandwidth) and 0.02 rad, and standard deviation less than 0.002 Hz and 0.5 rad for $f(t_5)$ and $\phi(t_5)$, respectively. This conclusion can also be drawn from the histograms on the second and fourth panels in Fig.~\eqref{fig:Pr(x|Y)}, the  shapes of which, by definition,   resemble the true posterior distributions $\Pr(f(t_5)|\mathbf{y})$ and $\Pr(\phi(t_5)|\mathbf{y})$.
\begin{figure}
    \includegraphics[width=1\linewidth]{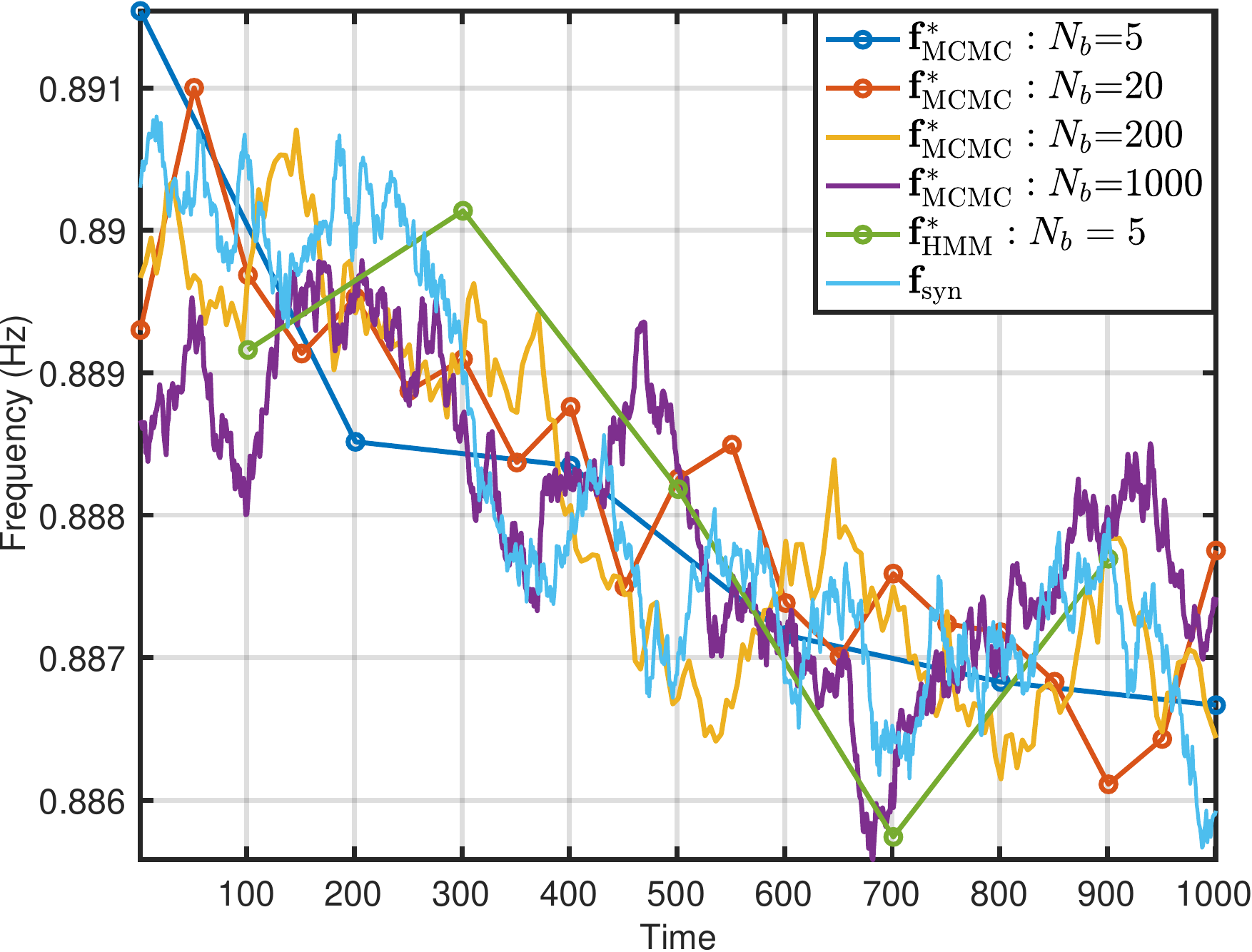}
    \caption{MAP frequency estimate ${\mathbf{f}^*_{\rm MCMC}}$ for $N_b=5\text{ (blue), }20\text{ (red)}$, $200$ (yellow) and $1000$ (purple) of the MCMC and $N_b=5$ (green) of the HMM at SNR = 0.15. The wandering dynamics of the frequency is captured even with $N_b=5$ knots.}
    \label{fig:MAP}
\end{figure}
 \subsubsection{ MCMC-MAP estimator}
     A typical realization of the  MCMC-MAP estimates of frequency paths $\mathbf{f}^*_{\rm MCMC}$ for $N_b=5,\ 20,\ 200$ and $1000$, compared with the HMM estimated frequency path is displayed in Fig.~\eqref{fig:MAP}. Here we can see, that the dynamics of the wandering frequency is captured even by $N_b=5$ knots. 
     
\begin{figure}
    \centering
    \includegraphics[width=1\linewidth]{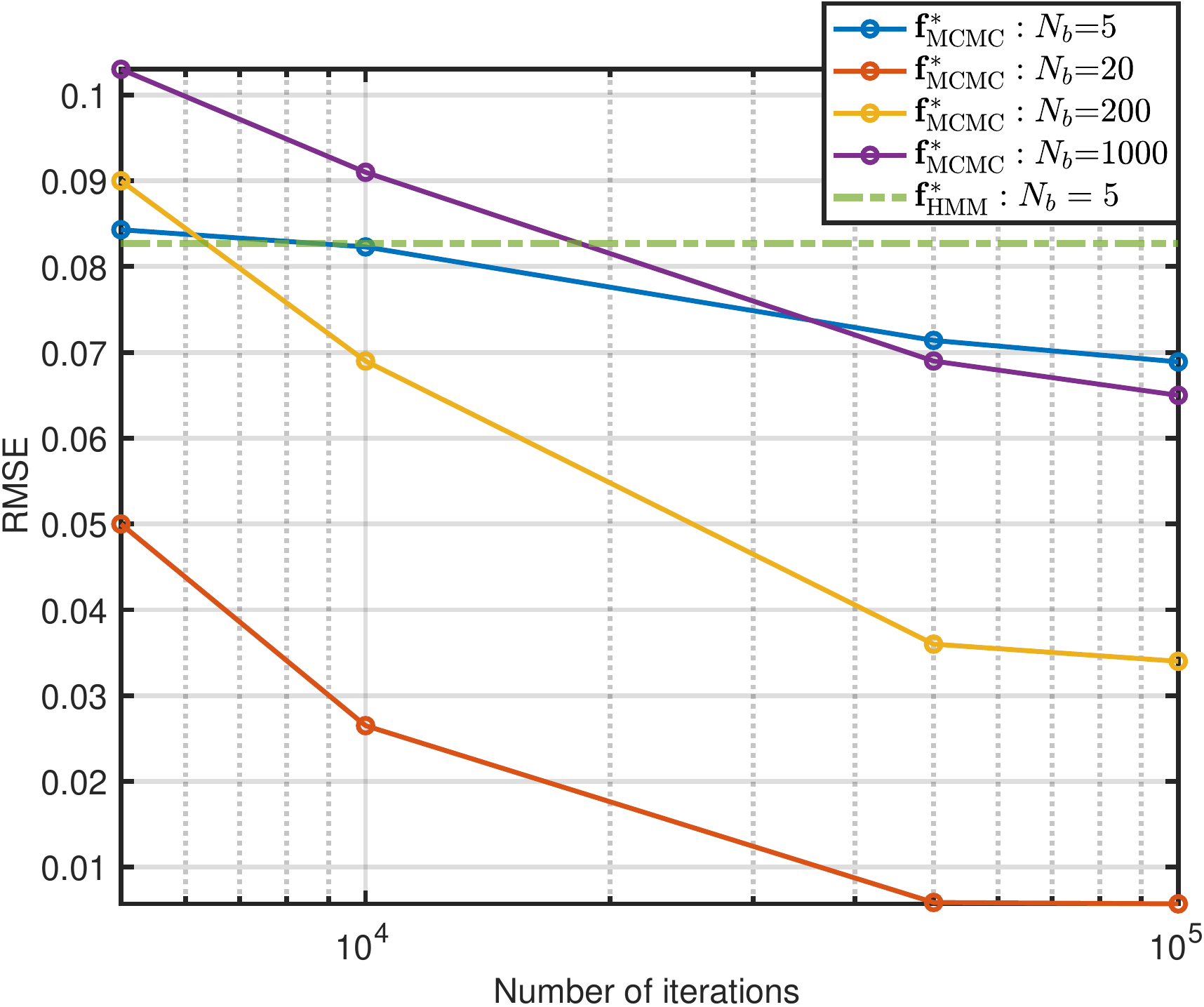}
    \caption{ RMSE of MCMC estimate as a function of number of blocks and number of iterations for $N_b=5\text{ (blue), }20\text{ (red)}$, $200$ (yellow) and $1000$ (purple)  and $N_b=5$ (green) of HMM at SNR = 0.15. Among all, MCMC with $N_b=20$ achieves the lowest mean error.}
    \label{fig:rmse}
\end{figure}

  In Fig.~\eqref{fig:rmse},  we plot the root-mean-square-error (RMSE) of the  MCMC-MAP estimated frequency path, normalized with respect to the path length $N$,
defined to be RMSE $\overset{\triangle}{=}\sqrt{E(||\mathbf{f}^*_{\rm MCMC}-\mathbf{f}_{\rm syn}||^2)/N}$, where $E(\cdot)$ denotes the sample  mean over $10^3$ experiments, and $\mathbf{f}^*_{\rm MCMC}$ and $\mathbf{f}_{\rm syn}$ denote the MCMC-MAP estimated and synthetic frequency path, respectively. 
In this example, every frequency point
$f^*_{\rm MCMC}(t_n)$ and $f{\rm syn}(t_n)$ takes values in the interval $[0,\;1]$, giving the upper bound for  RMSE of $1$. As shown here, $N_b=20$, among all, returns the lowest mean error and overall the MCMC-MAP estimator provides more accurate estimation against the HMM estimator, although at the cost of longer computing time.

\subsection{Detection performance}
\subsubsection{ MCMC-posterior distribution for detection: $\Pr(k|\mathbf{y})$}
\begin{figure}[h!]
    \includegraphics[width=\linewidth]{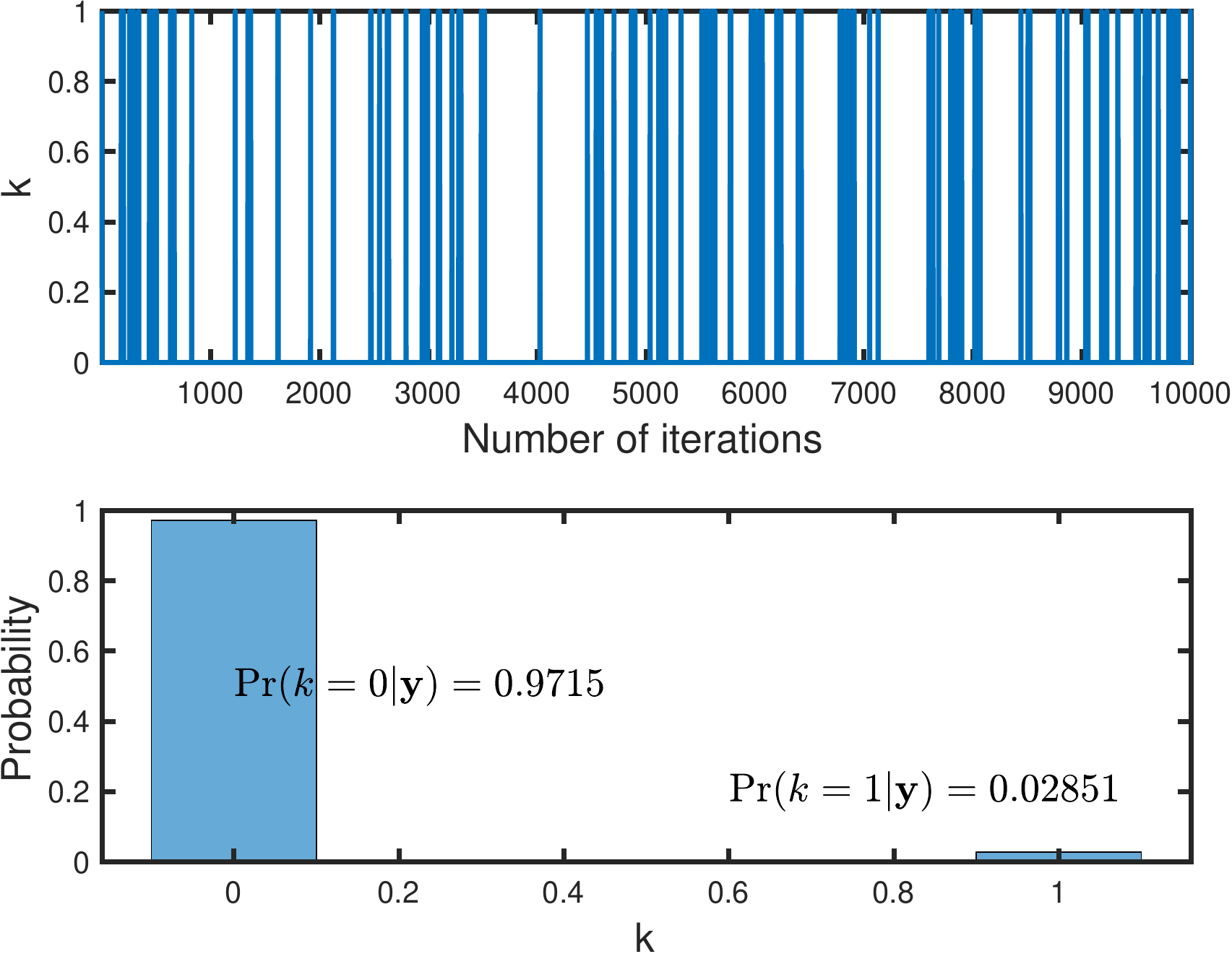}
    \includegraphics[width=\linewidth]{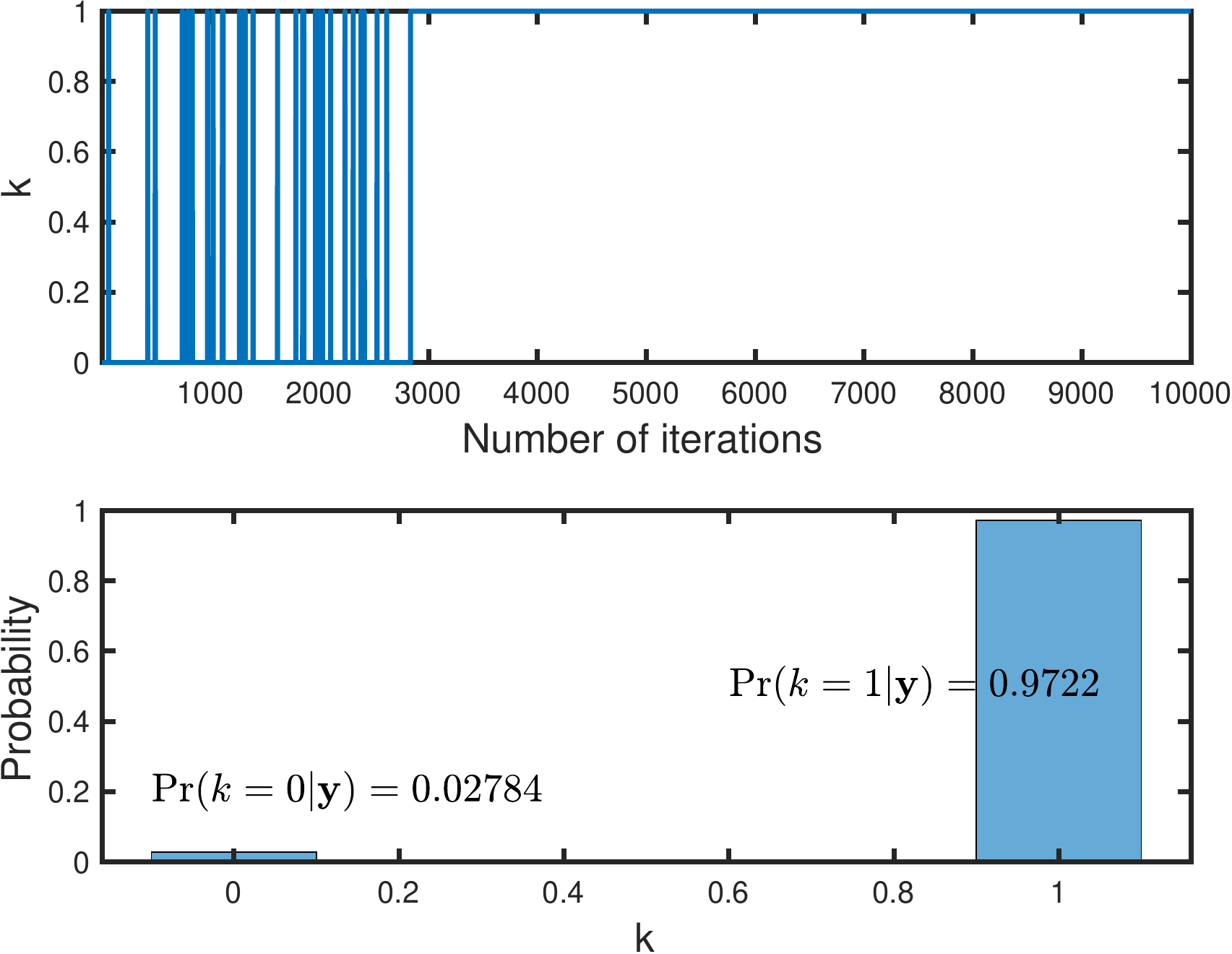}
\caption{Trace plot and histogram ( posterior distribution) of $k$. Upper panel:  Trace plot  of $k$ with synthetic data $k=0$. Second panel: Posterior distribution $\Pr(k=0|\mathbf{y})$ with synthetic data $k=0$. Third panel: Trace plot  of $k$ with synthetic data $k=1$ and $\text{SNR}=0.15$.  Lower panel: Posterior distribution $\Pr(k=1|\mathbf{y})$ with synthetic data $k=1$ and $\text{SNR}=~0.15$.}.
\label{fig:Pr(k|Y)}
\end{figure}

In Fig.~\eqref{fig:Pr(k|Y)},   two examples of the  MCMC-posterior distribution $\Pr(k|\mathbf{y})$ for an  $\text{SNR} = 0.15$ are presented.
 The upper panel shows a typical   trace plot of the parameter $k$   for when the data contain no signal, where the value of $k$ jumps constantly between $k=0$ and $k=1$.
 The histogram of this  $k$  is shown in the second panel. 
 The third panel shows typical  samples  when the signal is present; after around $3\times 10^3$ iterations, $k$  clearly approaches the  value~1. The lower panel depicts the histogram of $k$ for this case.
\subsubsection{Receiver operating characteristic}
\begin{figure}
    \includegraphics[width=1\linewidth]{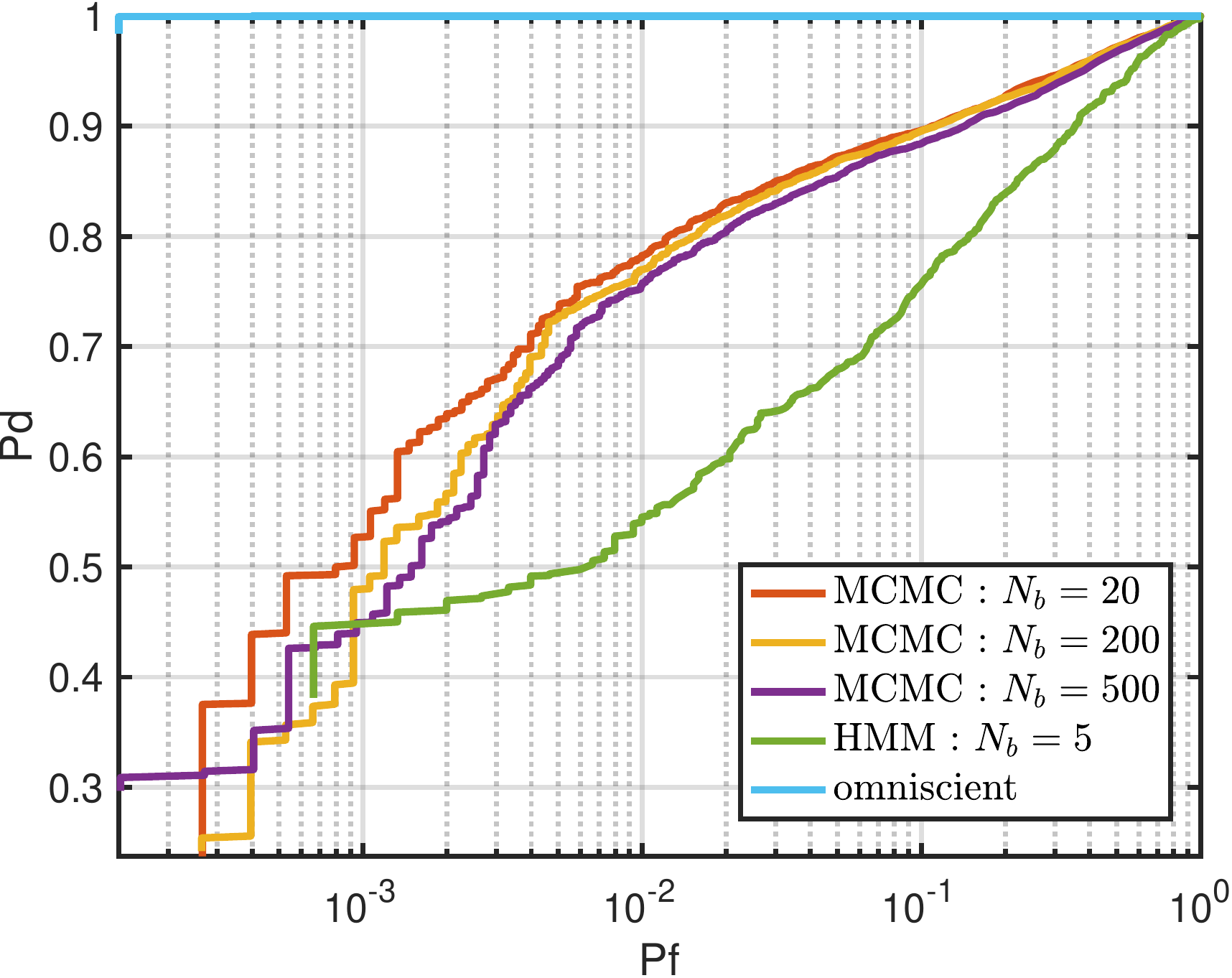}
    \includegraphics[width=1\linewidth]{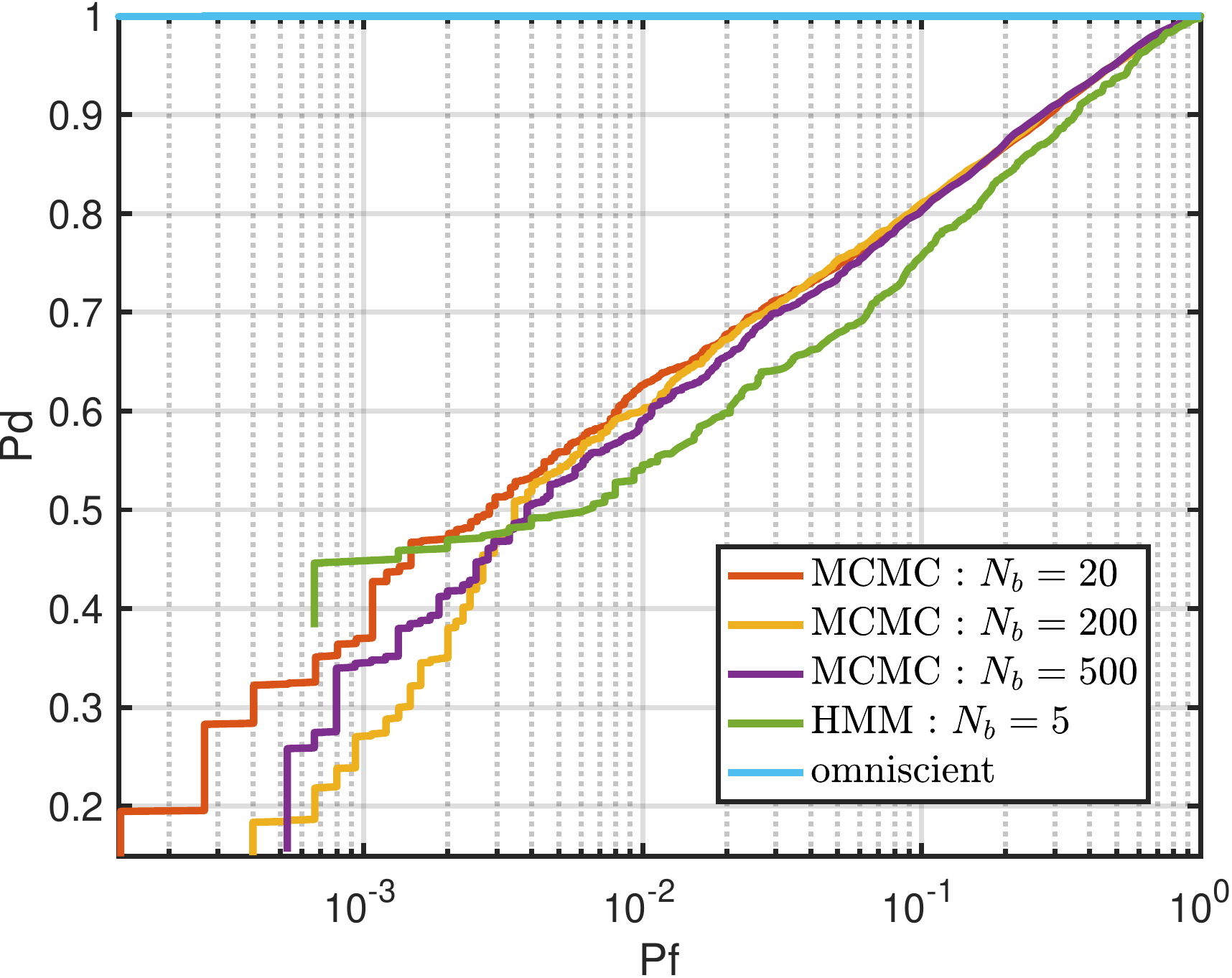}
    \caption{ROC curves at SNR = 0.15 for MCMC with $N_b=20\text{ (red), }200\text{ (yellow), }500\text{ (purple)}$ and HMM with $N_b=5$ (green). The upper panel shows ROC curves for $\gamma=10^{-4}$~Hz~sec$^{-1/2}$, and lower panel for $\gamma=10^{-5}$~Hz~sec$^{-1/2}$. The HMM detector yields worse performance when Pf $>10^{-3}$. The  omniscient detector (blue curve) provides an  upper bound.}
    \label{fig:roc1}
\end{figure}

\begin{figure}
    \centering
    \begin{minipage}{.45\textwidth}
    \includegraphics[width=1\linewidth]{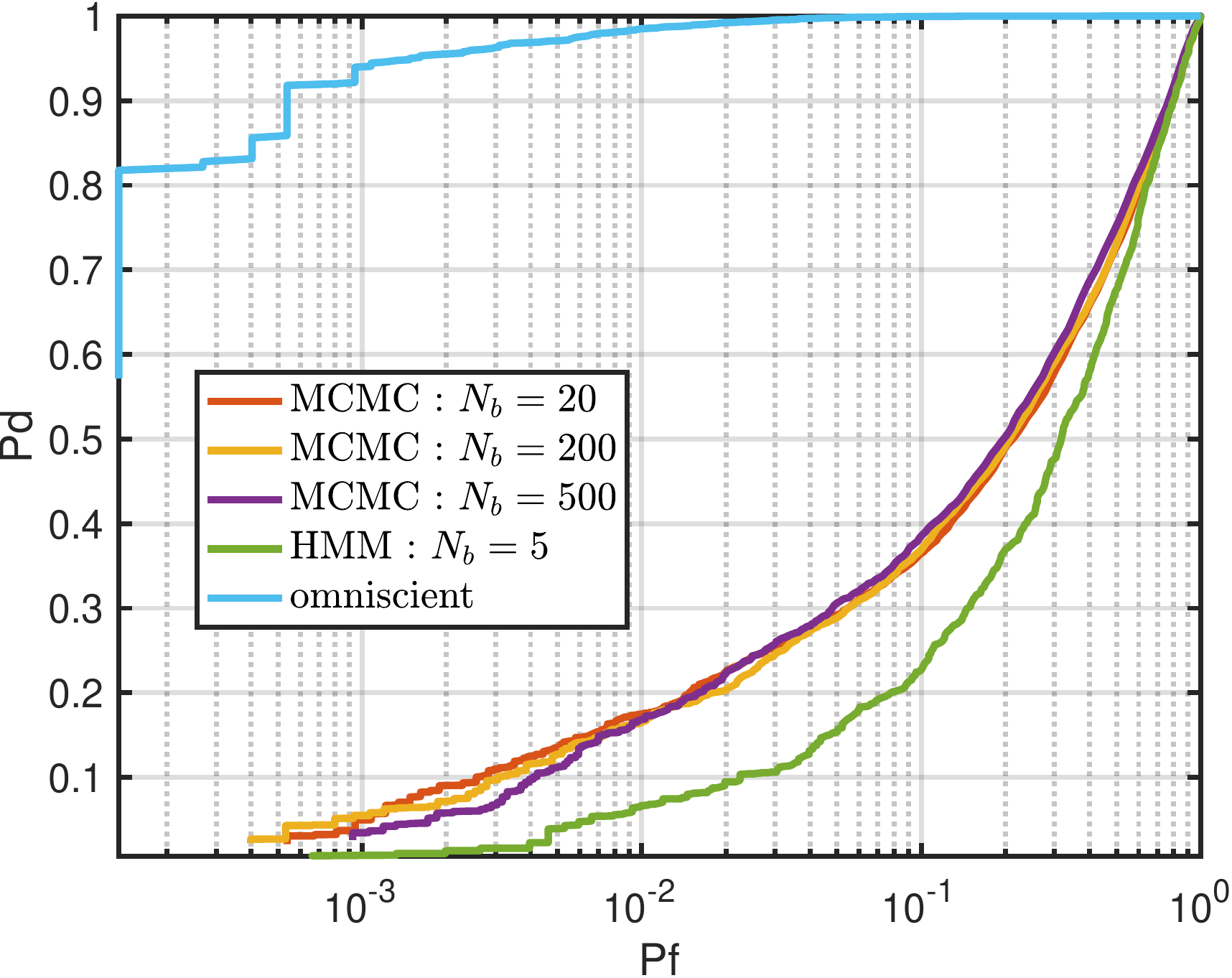}
    \end{minipage}
    \begin{minipage}{.45\textwidth}
    \centering
        \includegraphics[width=1\linewidth]{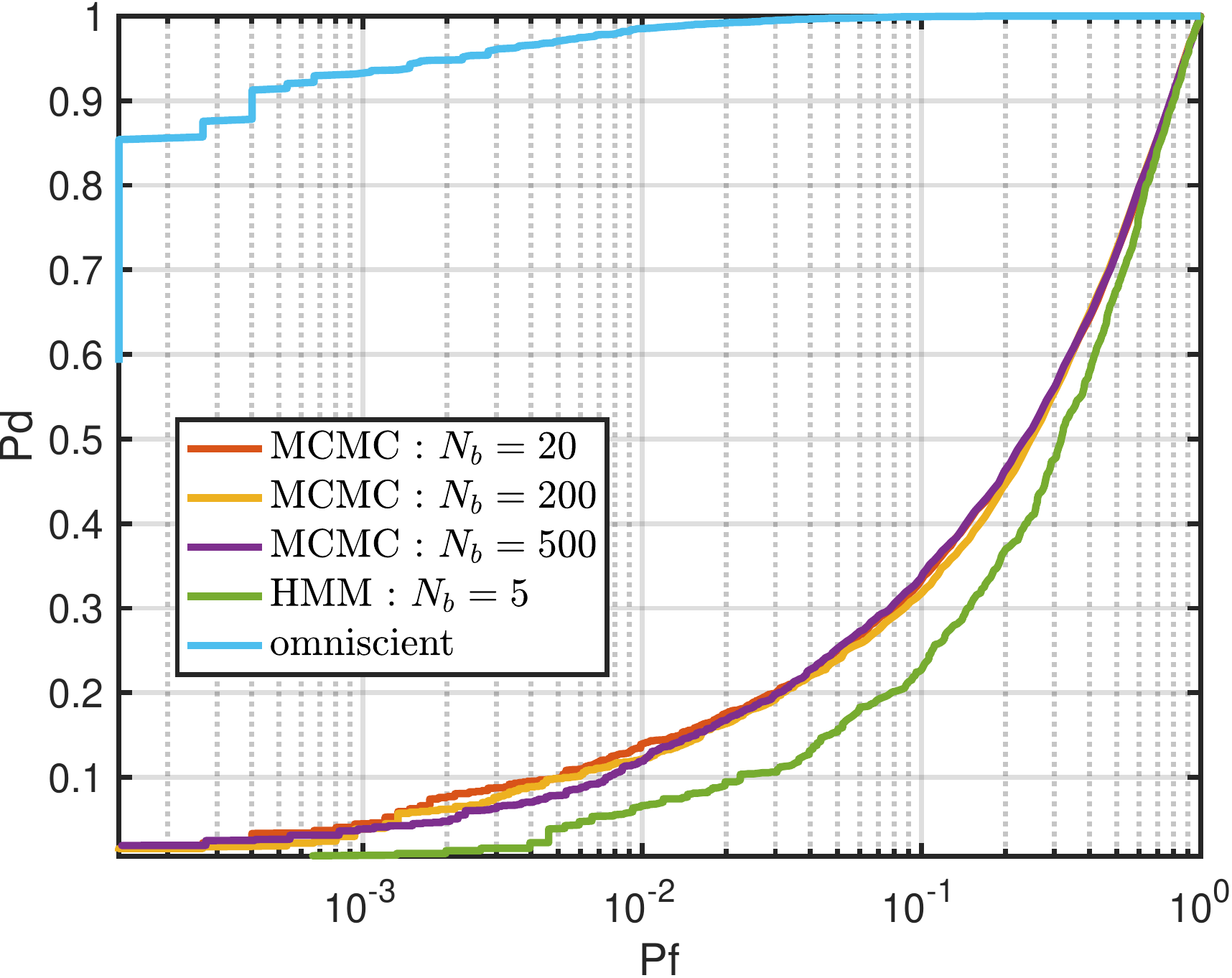}
        \end{minipage}
    \caption{ROC curves at SNR = 0.1 for MCMC with $N_b=20\text{ (red), }200\text{ (yellow), }500\text{ (purple)}$ and HMM with $N_b=5$ (green). The upper panel shows ROC curves for $\gamma=10^{-4}$~Hz~sec$^{-1/2}$, and lower panel for $\gamma=10^{-5}$~Hz~sec$^{-1/2}$. The  HMM detector yields the worst performance. The omniscient detector (blue) provides an  upper bound.}
    \label{fig:roc2}
\end{figure}

\begin{figure}
    \centering
    \begin{minipage}{.45\textwidth}
    \centering
    \includegraphics[width=1\linewidth]{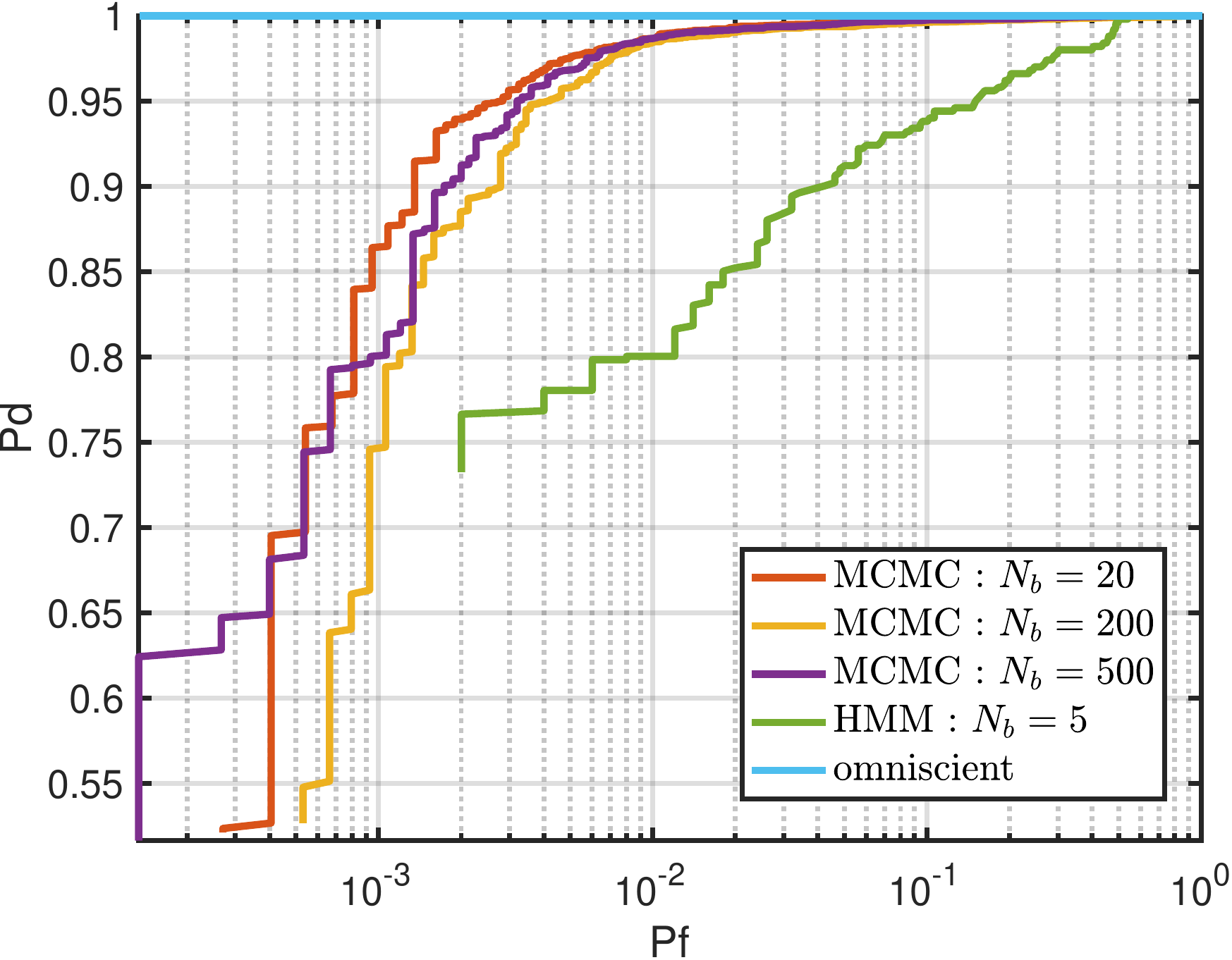}
\end{minipage}
    \begin{minipage}{.45\textwidth}
    \centering
        \includegraphics[width=1\linewidth]{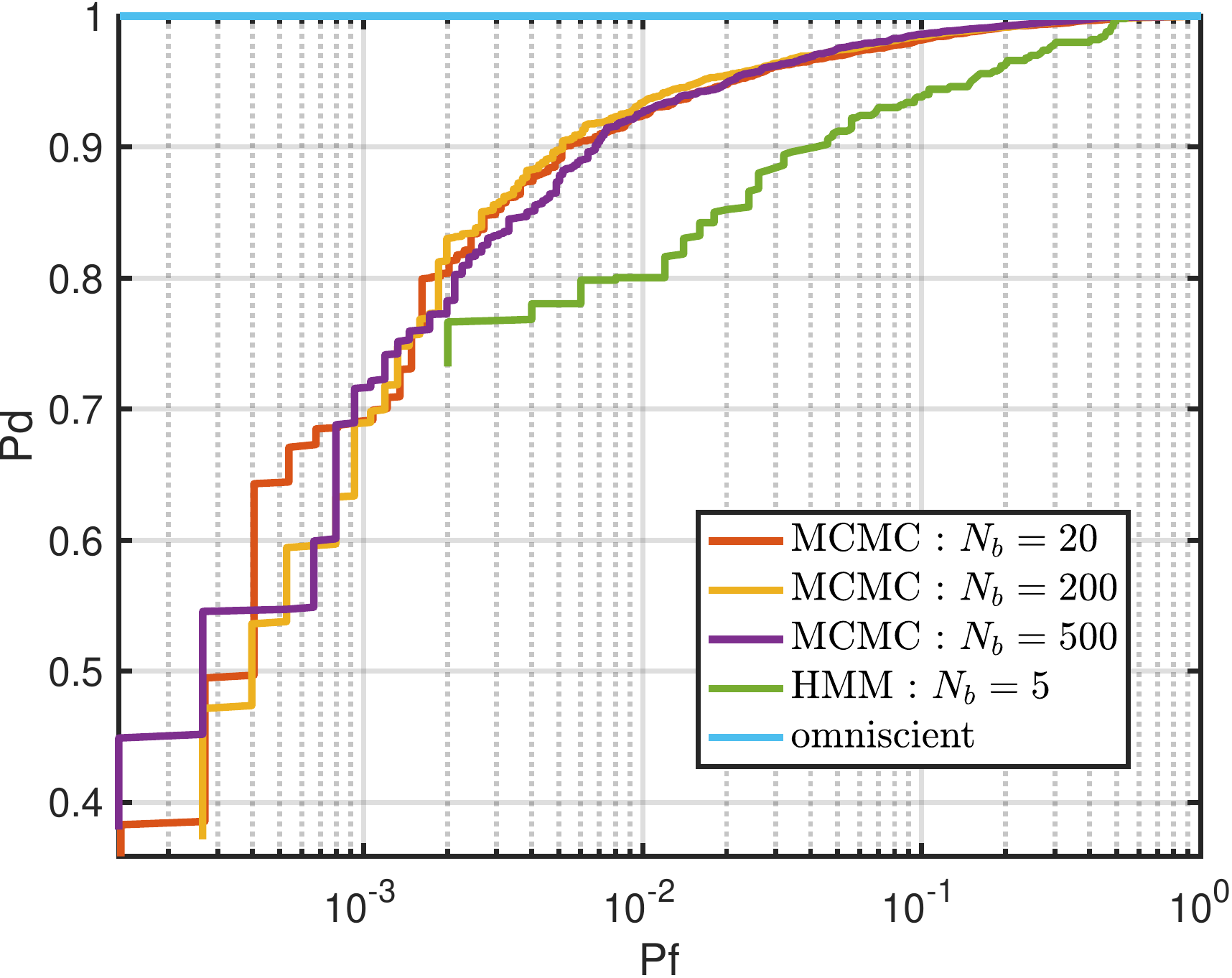}
        \end{minipage}
    \caption{ROC curves at SNR = 0.2 for MCMC with $N_b=20\text{ (red), }200\text{ (yellow), }500\text{ (purple)}$ and HMM with $N_b=5$ (green). The upper panel shows ROC curves for $\gamma=10^{-4}$~Hz~sec$^{-1/2}$, and lower panel for $\gamma=10^{-5}$~Hz~sec$^{-1/2}$. Overall, the  MCMC detector has overall better performance. The omniscient detector (blue)  provides an  upper bound.}
    \label{fig:roc3}
\end{figure}

 Receiver operating characteristic (ROC) curves for an omniscient,\footnote{The omniscient detector is based on the assumption that  the true  path $\mathbf{x}_{\rm {syn}}$ is known. It provides an upper bound for the probability of detection.}  The MCMC detector and the  HMM detector
 are shown in Fig.~\eqref{fig:roc1},~\eqref{fig:roc2} and~\eqref{fig:roc3},
computed over $10^5$ simulation runs at $\text{SNR} = 0.15$, 0.1 and 0.2 respectively for synthetic signals with  frequencies wandering according to \eqref{origin model} and \eqref{eq3}, with $\gamma_{\rm syn}=10^{-4}$ ~Hz~sec$^{-1/2}$.

The upper panels are for the MCMC algorithm with parameters $\gamma=10^{-4}$~Hz~sec$^{-1/2}$, and the lower  panels
are for $\gamma=10^{-5}$~Hz~sec$^{-1/2}$. The mismatch in $\gamma$ and $\gamma_\text{syn}$ appears to cause  degradation in the  MCMC detector performance. This sensitivity to  $\gamma$ is  an unwanted effect and requires further investigation.

At relatively high $\text{SNR} = 0.2$, the plots show that the MCMC detector outperforms the HMM detector across the whole Pf range.
 For $\text{SNR} = 0.1$, the detection rate for the MCMC detector, although higher than the HMM detector, is quite low, i.e.,  around 0.17 at Pf $=10^{-2}$.
At SNR = 0.15
both the MCMC detector and the HMM detector demonstrate better performance than when the $\text{SNR} = 0.1$ with the MCMC outperforming the HMM.
In particular, for a false alarm probability Pf $=10^{-2}$, the detection probability Pd of the MCMC detector is around 0.8 with matched $\gamma$'s, while dropping below 0.7 with mismatched $\gamma$'s.
%
In Fig.~\eqref{fig:roc1}, when $\text{SNR} = 0.15$, the MCMC detector outperforms the HMM detector across the Pf range greater than $10^{-2}$  for all choices of $N_b$ and $\gamma$.
%
  The HMM performs ``detection after estimation", i.e., it calculates the most likely frequency path first, then compares the statistics of this path with the statistics of the noise, while the detection is directly embedded in the design of the MCMC detector.
  As a result, the HMM's detection performance is heavily  dependent on the accuracy of estimation, as opposed to the MCMC detector, where estimation becomes a consequence of detection.
The degradation of performance at low SNR, known as the ``threshold effect''  is a common problem in nonlinear estimation. Even though we are not able to derive it mathematically, we infer from the plots that the  threshold effect for MCMC detector happens between $\text{SNR} = 0.15$ and $\text{SNR} = 0.1$.

Fig.~\eqref{fig:roc1},~\eqref{fig:roc2} and~\eqref{fig:roc3} also show that  $N_b$ has little effect on the overall detection performance of the MCMC detector. The red, yellow and purple curves  overlap each other, especially when 
Pf $<10^{-1}$. 

\begin{figure}
    \centering
    \begin{minipage}{.45\textwidth}
    \includegraphics[width=1\linewidth]{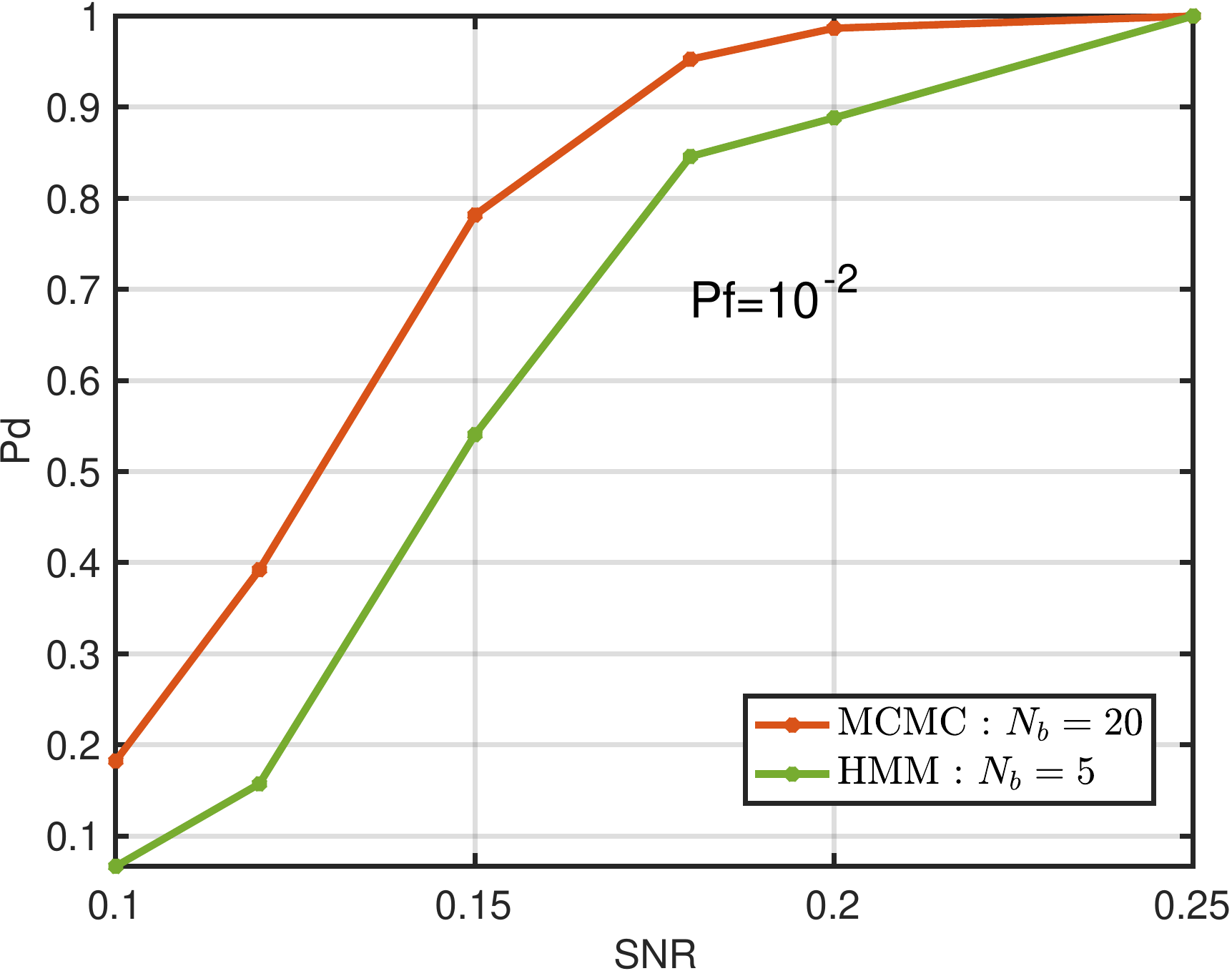}
    \end{minipage}
    \begin{minipage}{.45\textwidth}
        \centering
        \includegraphics[width=1\linewidth]{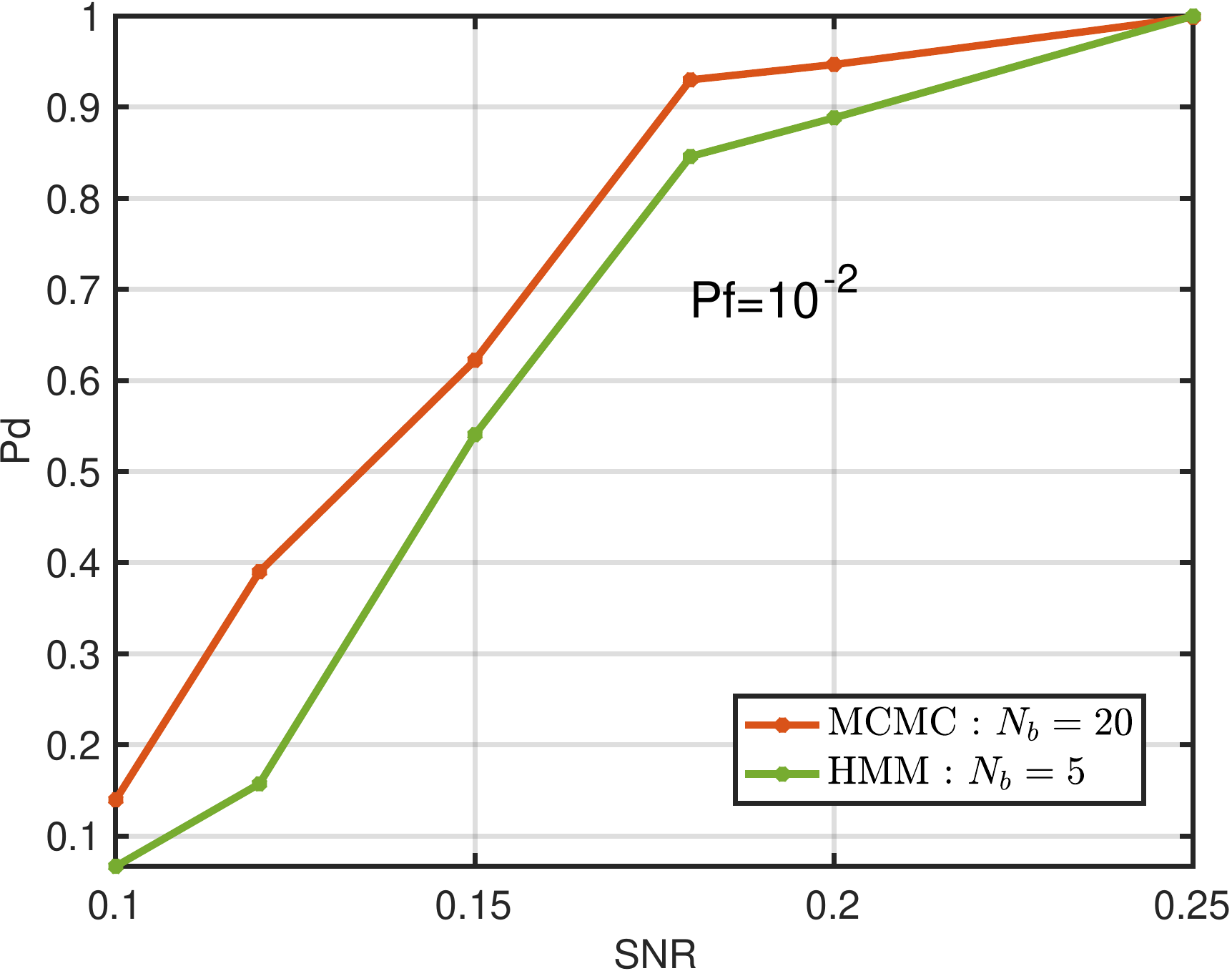}
    \end{minipage}
    \caption{Detection probability Pd versus SNR ranging from 0.1 to 0.25, with false alarm rate Pf = $10^{-2}$ for MCMC with $N_b=20$ (red) and HMM with $N_b=5$ (green). The upper panel shows ROC curves for $\gamma=10^{-4}$~Hz~sec$^{-1/2}$, and lower panel for $\gamma=10^{-5}$~Hz~sec$^{-1/2}$.  MCMC has higher detection probability than HMM across the SNR range regardless of the bias in $\gamma$.  }
    \label{roc4}
\end{figure}

In Fig.~\eqref{roc4}, we fix the  false alarm probability Pf = $10^{-2}$ and plot Pd versus SNR varying from 0.1 to 0.25 for the MCMC detector with $N_b=20$ and the HMM detector respectively.
Controlling the false alarm probability to be no more than $10^{-2}$ is typically tolerated in gravitational wave astrophysics applications \cite{2017PhRvD..95l2003A}.
 Similarly, the upper panel and lower panels show the MCMC detector's performance  without and with  mismatch in $\gamma$, respectively. In both plots the MCMC detector has higher detection probability than the HMM detector, even with $\gamma$ mismatched. For example when the SNR = 0.15, the MCMC detector outperforms the HMM detector with 25\% higher detection probability.



\section{Conclusion}
In this work a Bayesian  posterior density for detecting sinusoidal signals with wandering frequency in noise is derived and computed. The method is based on MCMC techniques.
As part of the algorithm, our method provides computation of the posterior density of the signal parameters. 
For efficient computation of this density we propose a knot-interpolating technique, where  we sample the signal parameters at the coarsely spaced time knots, while the rest of the signal is recovered by the interpolation between the knots.
A procedure  for selecting a reasonable number $N_b$ of knots, given the signal dynamics is presented and justified. This procedure
relies on the computation of  the (von~Neumann) entropy of the dynamics matrices.
 Although we cannot claim its optimality, we illustrate by experiments how the procedure provides a balance between the  runtime and  detection and estimation accuracy.

In addition, we have  developed an algorithm within MCMC  for proposing new  state paths that are arbitrarily close to the previous path. This method ensures  dense selection of MCMC samples for highly structured multi-dimensional vectors. The full description of the algorithm is provided.

The performance of the MCMC is evaluated in terms of  mean estimation errors and ROC curves  and compared with the performance of the HMM-based Viterbi algorithm. We demonstrate that our algorithm  presents both higher detection rates and greater estimation accuracy in all of the  experiments conducted.
In particular, the  simulation results show that  our method outperforms  the HMM  in  estimation accuracy by around 5\% and  improves  detection rate by up to 25\%.



%
%


 \bibliographystyle{ieeetr}

%
\section*{Appendix A: Proof of \eqref{eq3}}
\label{Appendix: derive covariance}
We consider a more general process where continuous frequency and phase path are described by
\begin{subequations}
\begin{align}
    df(t)&=-\frac{1}{\tau}f(t)dt+\gamma dB(t)\\
    \phi(t)&=\int_0^t f(s)ds
    \end{align}
\end{subequations}
where $\tau$ and $\gamma$ are called the  \emph{relaxation time} and the \emph{diffusion constant}.  

The variance and covariance of the time sampled variable $f(t+T)$ and $\phi(t+T)$ at  time $t+T$, with time increments $T>0$ are derived in \cite{suvorova2018phase} to be
\begin{subequations} \label{eq10}
\begin{align}
    \sigma_1^2\overset{\triangle}{=}&\text{ var}\{f(t+T)\}=(\gamma^2\tau/2)(1-e^{-2T/\tau})\\
    \sigma_2^2\overset{\triangle}{=}&\text{ var}\{\phi(t+T)\}=\gamma^2\tau^3[T/\tau-2(1-e^{-T t/\tau})\notag\\
    &+\frac{1}{2}(1-e^{-2T t/\tau})]\\
    \mathbf{K}\overset{\triangle}{=}&\text{ cov}\{f(t+T),\phi(t+T)\}\notag\\
    &=(\gamma^2\tau^2/2)(1-2e^{-T/\tau}+e^{-2T/\tau})
\end{align}
\end{subequations}
We consider $f(t)$ to be a Wiener process with diffusion constant $\gamma$ and $\tau= \infty$. Equation~\eqref{eq10} is then approximated by
\begin{subequations}
\begin{align}
    \sigma_1^2 &\approx \sigma^2T \\
    \sigma_2^2 &\approx \sigma^2{T}^3/3\\
    \mathbf{K} &\approx \sigma^2{T}^2/2
\end{align}
\end{subequations}
and we obtain the covariance matrix \eqref{eq3}.
\section*{Appendix B: pseudocode}
\begin{algorithm}
\KwIn{$\Big(\mathbf{x}(t_{n_1}),\mathbf{x}(t_{n_2})\Big)$.}
\KwOut{$\{\tilde{\mathbf{x}}(t_{n_1}),\tilde{\mathbf{x}}(t_{n_1+1}),\tilde{\mathbf{x}}(t_{n_1+2}),\dots,\tilde{\mathbf{x}}(t_{n_2})\}$.}
\begin{equation}
\Bigl\{
\begin{array}{ll}
\tilde{\phi}(t_{n_1+\ell})=\phi(t_{n_1})+f(t_{n_1})T\ell
+\frac{1}{2}b_1 (T\ell)^2+\frac{1}{3}b_2(T\ell)^3&\\
\tilde{f}(t_{n_1+\ell})=f(t_{n_1})+b_1T\ell+b_2(T\ell)^2, &
\end{array}
\end{equation}
with 
\begin{equation}
    \begin{bmatrix}
    b_1\\
    b_2
    \end{bmatrix}=
    \begin{bmatrix}
    T_b^2/2&T_b^3/3\\
    T_b & T_b^2
    \end{bmatrix}^{-1}
    \begin{bmatrix}
    {\phi}(t_{n_2})-\phi(t_{n_1})-f(t_{n_1})T_b\\
    {f}(t_{n_2})-f(t_{n_1})
    \end{bmatrix}
\end{equation}
for $\ell=0,1,\ldots,n_2-n_1$ with $T_b=t_{n_2}-t_{n_1}$.
\caption{function "Interp" }
\label{interp}
\end{algorithm}
\begin{algorithm}
\label{algorithm2}
Initialization: $k^1 \sim Bernoulli (1-\alpha),\tilde{\mathbf{x}}(t_1) =[\tilde{f}(t_1),\tilde{\phi}(t_1)],\Pr(s^1)=[\alpha \ 1-\alpha]^T$\;
Sample $u_1 \sim \mathcal{U}(0,1),u_2 \sim \mathcal{U}(0,1)$\;
  \uIf{$u_1< \alpha$}{$k^1 =0,\tilde{f}(t_1)=\varnothing,\tilde{\phi}(t_1) = \varnothing$, where the symbol $\varnothing$ indicates that $\tilde{\mathbf{x}}^i$ is meaningless for $k=0$}
  \Else{$k^1 =1,\tilde{f}(t_1)\sim \mathcal{U}(0,U),\tilde{\phi}(t_1) =0$}
 \Begin{
  \For{$i = 2:N_{\text{iteration}}$}{
    Update $\Pr(s^{i}=0)=\mathbf{\Gamma}_{00}\Pr(s^{i-1}=0)+\mathbf{\Gamma}_{10}\Pr(s^{i-1}=1)$\;
    \uIf{$\Pr(s^i=0)>u_2$}{$s^i=1$\;}
    \Else{$s^i=0$\;}
    \BlankLine
    Evaluate $s^i$ and $k^{i-1}$\;
    \uIf{$k^{i-1}=0$\bf{ and }$s^i=1$}{birth move: go to Algorithm 4\;}
    \uElseIf{$k^{i-1}=1$\bf{ and }$s^i=0$}{death move: go to Algorithm 5\;}
    \uElseIf{$k^{i-1}=1$\bf{ and }$s^i=1$}{update: go to Algorithm 6\;}
    \Else{assign $(\tilde{\mathbf{x}}^i,\tilde{a}^i,k^i)=(\varnothing,\varnothing,0)$.}
    }
    }
\caption{MCMC algorithm for joint detection estimation\label{joint}}
\end{algorithm}
\begin{algorithm}
\label{algorithm3}
\KwIn{$\tilde{\mathbf{x}}^{i-1},\tilde{a}^{i-1},k^{i-1}$.}
\KwOut{$\tilde{\mathbf{x}}^i,\tilde{a}^i,k^i$.}
    Propose a candidate state path according to Algorithm~6 (discussed in Section \ref{birth})\;
    Evaluate $q$ (Equation~\eqref{eq:q}) and $\bar{a}'$ based on $\tilde{\mathbf{x}}'$ obtained from Step 1 (Equation~\eqref{eq:abar})\;
    Sample $\tilde{a}'$ from distribution $\tilde{a}' \sim \mathcal{CN}(0,\sigma^2q)$;
    or (optional) let $\tilde{a}'=\bar{a}$ for simplicity if $\tilde{a} $ is of little interest to us\footnotemark\;
    Evaluate $W_{\tilde{a}}'$ and $W_f'$ based on $\tilde{a}'$ and $\tilde{\mathbf{x}}'$ (Equation~\eqref{Wf}-\eqref{eq:wa})\;
    Accept $(\tilde{\mathbf{x}}^i,\tilde{a}^i,k^i)=(\tilde{\mathbf{x}}',\tilde{a}',1)$ with probability $A_{\text{birth}}$ (Equation~\eqref{abirth}); otherwise set $(\tilde{\mathbf{x}}^i,\tilde{a}^i,k^i)=(\tilde{\mathbf{x}}^{i-1},\tilde{a}^{i-1},k^{i-1})$.
\caption{``Birth'' move}
\end{algorithm}
\footnotetext{In our MCMC algorithms, we do not sample $\tilde{a}$ to evaluate \eqref{wtildea}, but only compute  the MAP  estimate of $\tilde{a}$ for a realization of $\mathbf{D}_f$, which is $\bar{a}$  and we simply set the amplitude proposals $\tilde{a}'=\bar{a}$.}
\begin{algorithm}
\label{algorithm4}
\KwIn{$\tilde{\mathbf{x}}^{i-1},\tilde{a}^{i-1},k^{i-1}$.}
\KwOut{$\tilde{\mathbf{x}}^i,\tilde{a}^i,k^i$.}
    Given previous value $\tilde{\mathbf{x}}^{i-1}\text{ and } \tilde{a}^{i-1}$ ($k^{i-1}$ must be 1)\;
    Evaluate $W_{\tilde{a}}^{i-1}$ and $W_f^{i-1}$ based on $\tilde{\mathbf{x}}^{i-1};\ \tilde{a}^{i-1}$ (Equation~\eqref{Wf}-\eqref{eq:wa})\;
    Accept $(\tilde{\mathbf{x}}^i,\tilde{a}^i,k^i)=(\varnothing,\varnothing,0)$ with probability $A_{\text{death}}$ (Equation~\eqref{adeath}); otherwise set $(\tilde{\mathbf{x}}^i,\tilde{a}^i,k^i)=(\tilde{\mathbf{x}}^{i-1},\tilde{a}^{i-1},k^{i-1})$.
\caption{``Death Move''}
\end{algorithm}
\begin{algorithm}
\label{algorithm5}
\KwIn{$\tilde{\mathbf{x}}^{i-1},\tilde{a}^{i-1},k^{i-1}$.}
\KwOut{$\tilde{\mathbf{x}}^i,\tilde{a}^i,k^i$.}
    Propose a candidate state path according to Algorithm~7 (described in Section~\ref{update})\;
    Evaluate $q$ (Equation~\eqref{eq:q}) and $\bar{a}'$ based on $\tilde{\mathbf{x}}'$ obtained from Step 1 (Equation~\eqref{eq:abar})\;
    Sample $\tilde{a}'$ from distribution $\tilde{a}' \sim \mathcal{CN}(0,\sigma^2q)$;
    or (optional) let $\tilde{a}'=\bar{a}$ for simplicity if $\tilde{a} $ is of little interest to us\;
    Evaluate $W_{\tilde{a}}'$ and $W_f'$ based on $\tilde{a}'$ and $\tilde{\mathbf{x}}'$ (Equation~\eqref{Wf}-\eqref{eq:wa})\;
    Accept $(\tilde{\mathbf{x}}^i,\tilde{a}^i,k^i)=(\tilde{\mathbf{x}}',\tilde{a}',1)$ with probability $A_{\text{update}}$ (Equation~\eqref{aupdate}); otherwise set $(\tilde{\mathbf{x}}^i,\tilde{a}^i,k^i)=(\tilde{\mathbf{x}}^{i-1},\tilde{a}^{i-1},k^{i-1})$.
\caption{``update'' move}
\end{algorithm}
\begin{algorithm}
\label{algorithm6}
\KwIn{$\varnothing$.}
\KwOut{$\tilde{\mathbf{x}}' $.  }
Initialization: $\tilde{\mathbf{x}}'(t_1)=\{\tilde{\phi}'(t_1),\tilde{f}'(t_1)\}$ with $\tilde{\phi}'(t_1)=0$, $\tilde{f}'(t_1) \sim \mathcal{U}(0,U)$\;
Place the knots, i.e. determine the value for $N_b$ and the corresponding $T_b,M$\;
Generate  knot positioned path $\mathbf{x}'_{N_b}$, i.e. start from $\mathbf{x}'_{N_b}(1)=\mathbf{\tilde{x}}'(t_1)$ in Step 1, and follow the sampled path model (Equation~\eqref{Reducedpath}), with random noise $\mathbf{w}(j) \overset{i.i.d.}{\sim} \mathcal{N}(\mathbf{ 0,C}) $ with $\mathbf{C}$ given in Equation~\eqref{eq:covmatrix}\;
Call function Interp$\Big(\mathbf{x}'_{N_b}(j)$, $\mathbf{x}'_{N_b}(j+1)\Big)$ for $j=1,\dots,N_b-1$ from Algorithm 1 and obtain the proposed sample path $\tilde{\mathbf{x}}'$.
\caption{Generate a new proposal path $\mathbf{x}'$ for the ``birth'' step}
\end{algorithm}
\begin{algorithm}
\KwIn{$\tilde{\mathbf{x}}^{i-1}$}
\KwOut{$\tilde{\mathbf{x}}'$. }
Extract $N_b$ knots along $\tilde{\mathbf{x}}^{i-1}$ and form $\mathbf{x}^{i-1}_{N_b}(m+1)=\tilde{\mathbf{x}}^{i-1}(t_{mM+1})$ for $m=0,\dots,N_b-1$\;
 Uniformly sample an integer $l \in \{1,2,\dots,N_b\}$, and determine the starting point $\mathbf{x}^{i-1}_{N_b}(l)$ to be the $l$th element of $\mathbf{x}_{N_b}^{i-1}$; choose a value for $\beta$\;
Determine the matrix $\mathbf{M}_1'$ and $\mathbf{M}_2'$ based on $l$ from Step 1 (Equation~\eqref{M_1'})\;
Get previous noise sequence $\mathbf{q}^{i-1}$, which satisfies $\mathbf{M}_2'\mathbf{q}^{i-1}=\mathbf{x}_{N_b}^{i-1}-\mathbf{M}_1'\mathbf{x}_{N_b}^{i-1}(l)$\;
Generate a noise sequence $\mathbf{q}'=[\mathbf{q}^T(0),\dots, \mathbf{q}^T(N_b-1)]^T$, where $\mathbf{q}^T_n  \overset{i.i.d.}{\sim} \mathcal{N}(\mathbf{0},\mathbf{I}_2)$\;
Compute new noise sequence $\mathbf{q}^{i}$  via $\mathbf{q}^{i}=\mathbf{q}^{i-1}\cos{\beta}+\mathbf{q}' \sin{\beta}$;
Generate proposed sample path $\mathbf{x}_{N_b}'$, where $\mathbf{x}_{N_b}'=\mathbf{M}_1'\mathbf{x}_{N_b}^{i-1}(l)+\mathbf{M}_2'\mathbf{q}^{i}$\;
Call function Interp$\Big(\mathbf{x}'_{N_b}(j)$, $\mathbf{x}'_{N_b}(j+1)\Big)$ for $j=1,\dots,N_b-1$ from Algorithm 1 and obtain the proposed sample path $\tilde{\mathbf{x}}'$.
 \caption{Generate a proposal path $\mathbf{x}'$ for the ``update'' step}
 \label{algorithm7}
\end{algorithm}
\end{document}